\documentclass[a4paper]{article}
\usepackage[english]{babel}
\usepackage{amsmath}
\usepackage{graphicx}
\usepackage{tabularx}
\usepackage{tipa}
\usepackage[colorinlistoftodos]{todonotes}
\usepackage{authblk}
\usepackage{placeins}
\usepackage{hyperref}
\usepackage{epigraph}
\hypersetup{
    colorlinks=true,
    linkcolor=blue,
    filecolor=magenta,      
    urlcolor=blue,
}

\title{On the Records}
\author{Santa Fe Institute Postdocs\footnote{This manuscript is a
    product of {\it 72 Hours of Science} (see Appendix
    \ref{statement}). Authorship is uniformly attributed to, in
    alphabetical order, Andrew Berdahl, Uttam Bhat, Vanessa Ferdinand,
    Joshua Garland, Keyan Ghazi-Zahedi, Justin Grana, Joshua A.\ Grochow, Elizabeth Hobson, Yoav Kallus, Christopher P.\ Kempes, Artemy Kolchinsky, Daniel B.\ Larremore, Eric Libby, Eleanor A.\ Power, and Brendan D.\ Tracey}}
\affil{Santa Fe Institute, Santa Fe, NM, 87501 USA}
\affil{benepinamics@gmail.com}
\date{}
\begin{document}
\maketitle


\begin{abstract}
World record setting has long attracted public interest and scientific investigation. Extremal records summarize the limits of the space explored by a process, and the historical progression of a record sheds light on the underlying dynamics of the process. Existing analyses of prediction, statistical properties, and ultimate limits of record progressions have focused on particular domains. However, a broad perspective on how record progressions vary across different spheres of activity needs further development. Here we employ cross-cutting metrics to compare records across a variety of domains, including sports, games, biological evolution, and technological development. We find that these domains exhibit characteristic statistical signatures in terms of rates of improvement, ``burstiness'' of record-breaking time series, and the acceleration of the record breaking process. Specifically, sports and games exhibit the slowest rate of improvement and a wide range of rates of ``burstiness.'' Technology improves at a much faster rate and, unlike other domains, tends to show acceleration in records. Many biological and technological processes are characterized by constant rates of improvement, showing less ``burstiness'' than sports and games. It is important to understand how these statistical properties of record progression emerge from the underlying dynamics. Towards this end, we conduct a detailed analysis of a particular record-setting event: elite marathon running. In this domain, we find that studying record-setting data alone can obscure many of the structural properties of the underlying process. The marathon study also illustrates how some of the standard statistical assumptions underlying record progression models may be inappropriate or commonly violated in real-world datasets.
\end{abstract}


\section{Introduction}

How deep can a person dive? How tall can a building be? To answer these questions, we might look at how deep a person has ever dived, or how tall a building has ever been. This is the world record for these particular achievements. A record is an extremal event exhibited by some stochastic or dynamical process. The sequence of record setting provides a summary statistic of the history of a process. Records may also shed light on the ultimate limit of a process, and for this reason, previous work has focused on the connection between records and the limits of human or animal performance (e.g. \cite{noubary_survival_2004,radicchi_universality_2012}). The temporal progressions of record may also help us understand the diversity of underlying mechanisms, since technological innovation, shifts in population, and the rarity of exceptional performances or individuals will affect the progression dynamics. These processes have been investigated using a variety of empirical and theoretical techniques.

Extremal events, including records, are a classically studied phenomena in both the physical and social sciences. A canonical example is maximal displacement in the time series of random walks (e.g. \cite{redner2001guide}). The statistical properties of record-breaking has drawn attention from the stochastic process community \cite{chandler_distribution_1952,aitken_statistical_2004,berthelot_athlete_2010,berthelot_citius_2008,charalambides_distribution_2007,einmahl_records_2008,gembris_sports_2002,gembris_evolution_2007,glick_breaking_1978,holmes1969note,kelley2006predicting,krug_records_2007,majumdar_universal_2008,majumdar_universal_2010,nagy_statistical_2013,nevzorov_records_1988,nevzorov_record_1998,noubary_survival_2004,noubary_tail_2010,noubary_procedure_2005,radicchi_universality_2012,redner_role_2006,solow_how_2005,terpstra_simple_2007,volf2010stochastic,wergen_record_2012,wergen2013records,yang_distribution_1975,franke2010records}. Generally, this work considers how the underlying properties of a stochastic process relate to properties of the resulting derived process of record-setting events (time between records, for example). These theoretical results have been applied extensively to real sports data, as have other empirical investigations on record progression in swimming, track and field events, sprinting and distance running \cite{katz_power_1999,lippi_updates_2008, nevill_are_2005, nevill_are_2007,berthelot_citius_2008,berthelot_athlete_2010}. 

Beyond sports, other work has considered records in temperature and climate data \cite{redner_role_2006,coumou2012decade} as well as technological innovation \cite{moore_cramming_1965,bettencourt_determinants_2013,koh_functional_2006,nagy_statistical_2013,mcnerney_role_2011,nagy_superexponential_2011, lundstrom_moores_2003}. Perhaps the most famous of the technological examples is Moore's law, which describes the advances in computer processor capacity in time \cite{moore_cramming_1965}. Other work has considered record-setting in technological advances and improvements in production curves \cite{bettencourt_determinants_2013,koh_functional_2006,nagy_statistical_2013,mcnerney_role_2011,nagy_superexponential_2011, lundstrom_moores_2003}. 

Although any time-series can be used to generate a progression of records, the mechanisms, dynamics, and limits of each system may be radically different between phenomena. For example, a common progression seen in many records, such as horse racing, is characterized by a phase of increase followed by an apparent saturation, in which new records may be explained by the sampling of rare events from a stationary distribution \cite{denny_limits_2008,berthelot_athlete_2010,desgorces_oxford_2008,prampero_factors_2003,nevill_are_2005,lundstrom_moores_2003,chang2011limit}. Other processes, such as Moore's law, were for a long time characterized primarily by rapid expansion before hitting emerging limits \cite{moore_cramming_1965,lundstrom_moores_2003}. Thus, some processes may be  characterized by a growth phase, others by the saturated dynamics, and some by a transition between the two. 

Many past efforts have addressed the underlying mechanisms of these dynamics. However, most of this previous work has focused on records in specific domains (e.g. sports), or employ models with strong statistical or mechanistic assumptions (e.g., model records as arising from independent and identically distributed [IID] processes or random walks). Consequently, there is a need to understand both the entire space of mechanisms and how the characteristics of record progressions vary across disparate domains.

In this paper, we develop cross-cutting metrics for comparing records in a variety of domains, including sports, games, biological evolution, and technological improvement. Our main goal is to use information about record setting to develop a domain-agnostic characterization of the processes that drives the record setting dynamics. Given these goals, we (1) gather a wide variety of record data across domains, (2) develop metrics for contrasting these processes, and (3) compare the spectrum of record phenomena to different detailed processes and mechanisms.  We compare and contrast two main categories of record progression: Record progressions which (presumably) approach and are constrained by fundamental limits, such as 100 meter dash or other physical sports which encounter limits of human performance, and record progressions which have to do with the cumulative exploration of some underlying space, possibly with various constraints. This includes biological evolution and technological progression. We show that these different types have different statistical ``signatures.€''

In particular, we characterize record progressions in terms of:
\begin{enumerate}
\item Acceleration (or deceleration) of new records: Are records arriving more or less frequently?  Under IID assumptions, we would expect the latter \cite{krug_records_2007}. However, we see that in many domains, the arrival of records is speeding up.
\item Acceleration (or deceleration) of achieved values: are new records making incremental or substantial improvements over previous records?
\item Burstiness: Are records coming in clustered bursts? The ``burstiness''™ of record progression has been documented in systems ranging from road cycling to ski jumping to boat racing \cite{desgorces_oxford_2008, el_helou_tour_2010}.  Dramatic bursts have been connected to advances in technology, training, illicit activities (such as doping), or simply the natural dynamics of a particular system. Another possible reason for the burstiness of record settings is the effect of exceptional individuals, who set multiple records during a time-span over which they were active. For example, in the men's pole vault, Sergey Bubka set 19 records over a 10-year period.
\end{enumerate}

\section{Theoretical Background}
\label{sec:background}

Extremal events have undergone considerable study in the statistical literature.

Early treatments (e.g., \cite{neuts_waiting_1967,holmes1969note}) consider a model in which samples are drawn IID from a stationary probability distribution, keeping track of the maximum value observed under N samples. This analysis led to the Fisher-Tippet-Gnedenko theorem, and the Generalized Extreme Value distribution, which give the probability density for the distribution over the extreme values (see Ref.~\cite{wergen2013records}).

This IID sampling analysis was also used to examine the expected progression of records as time goes on. Using a permutation argument, it has been shown that the probability of the $n$\textsuperscript{th} sample being a record is $1/n$. This implies that the expected time for the next record, under IID sampling, is unbounded (see, e.\,g., \cite[p.~15]{feller1966book}). Less obviously, the indicator variable $I_n$ of the event that the $n$\textsuperscript{th} draw $x_n$ is a record is uncorrelated with $I_{n-1}$ (and any other previous indicator variables or combinations thereof). That is, under IID sampling, observing that a record has been set does not make it any more or less likely that a new record will be set going forward. In these simple models, it is also possible to analytically derive information about the distribution of the waiting times between successive record attempts; a theorem originally of Neuts \cite{neuts_waiting_1967} and strengthened by Holmes \& Strawderman \cite{holmes1969note} says that the distribution of the waiting time between the $n$\textsuperscript{th} record and the preceding record is exponential in $n$ almost surely. 

Based on actual record data, these results from IID processes do not seem sufficient to predict most record progressions (for example, the Men's 100m dash record was set in 2005, 2006, 2007, 2008 and 2009). This has lead researchers to explore other generative processes. Here we highlight a few key examples. See Wergen \cite{wergen2013records} and citations therein for a more thorough introduction. 

Perhaps the next simplest model, which captures more real-world phenomena, still assumes IID draws from a stationary distribution, but instead of 1 draw per timestep, there are $k$ draws per timestep and the maximum of these is reported, viz. $k$ people running a marathon in year $n$, with $x_n$ being the time of the winning person. Then the ``population'' $k$ is allowed to grow with time. This could reflect, for example, increasing worldwide population, increasing participation in a given competition, or some combination of the two. Yang \cite{yang_distribution_1975} studies this model where the population $k$ is exponentially growing, and observed that ``the rapid breaking of Olympic records is not due mainly to the increase in population.'' More specifically, if the population is growing with rate $\lambda$---that is, $k_t = \lambda^t k_0$---then, using $\Delta_n$ to denote the waiting time between the $n$\textsuperscript{th} record and the preceding record, Yang showed:
\[
\lim_{n \to \infty} Pr(\Delta_n = \delta) \propto \lambda^{-\delta}.
\]
In other words, in the limit of a large number of records, the waiting time in this model is distribution according to an exponential distribution with parameter $\lambda$. We test this as a null model on our datasets in Section~\ref{sec:waitingtime}.

Another common assumption is the Linear Drift Model (LDM), where there is a fixed underlying distribution with a linear added drift, viz. $X \sim Y +c t$, where $c$ is some drift speed and $Y$ is IID and time-invariant \cite{ballerini_resnick_1985_linear, ballerini_resnick_1987_linear, borovkov1999linear, franke2010records, WFK2011linear}. This model has seen particular employment with the Gaussian Linear Drift Model, most notably in the climate setting. In the climate, there is a background forcing of temperature due to global warming, which increases the set of observed temperatures above the otherwise chaotic distribution of daily weather events. Redner \textit{et al.} \cite{redner_role_2006}, and then others, use this model to analyse the expected changes in background forcing, and thus infer the change in global mean temperature through the observation of record events. A second common model is the increasing variance model (see, e.\,g., \cite{krug_records_2007, eilazar_klafter_2009_variance}) where $f_k(x_k) = \lambda_k f(\lambda_k x)$, where $\lambda_k$ is typically $k^\alpha$. Increasing population is a specific case of this model.

These two models capture different potential phenomena: linear drift attempts to capture improvement in the base rate of performers, while increasing variance could capture number of participants and variance of their performances. Yet, these two cases can be hard to distinguish from one another in real data, as they yield similar predictions for the progression of records. One notable difference between the two predictions arises in the correlations between records: In the case of linear drift, seeing a record broken makes it more likely a new record will be broken (positive correlation between successive waiting times). In contrast, in the case of increasing variance, successive waiting times are \textit{negatively} correlated: Seeing a record broken makes it less likely for any particular observation to be a new record. While this could be a useful metric to distinguish these two models, in practice it can be hard to operationalize, especially for time series where there are only record observations, but no observations of attempts that do not produce records.

While the literature on record progression of stochastic processes is substantial, the set of models with successful numeric treatment is still not very large. There is little categorization of the kinds of dynamics observed in record progressions in general, and comparison with empirical studies has focused on sports, or places where we have good characterization on the underlying non-stationarity in the distribution (for example the linear drift in climate, or Brownian motion in finance). 

\section{Classification of Records}

\subsection{Records Dataset}

Much of the work on records to date has focused on particular types of records, many of which appear to have similar patterns of progression. Record progressions in athletics, particularly Olympic records, have been well studied, and frequently suggest that a limit to human performance is being approached \cite{nevill_are_2005,berthelot_athlete_2010, radicchi_universality_2012}. In contrast, others have studied the progression of technological innovations, which often suggest something closer to exponential growth \cite{mcnerney_role_2011,nagy_superexponential_2011,nagy_statistical_2013}. To better understand the full phenomenological range of record progressions, we collected a wide range of record progressions from a wide range of domains with potentially disparate dynamics. In addition to the classic dataset of world records for Olympic events, we compiled data on other sports, including horse racing and competitive hotdog eating. For technological progressions, we drew from data used in Nagy et al.\ \cite{nagy_statistical_2013} on computational power, as well as records of other forms of technological advances, such as the tallest buildings, speed records for boats, largest pumpkins, and crop yield. We compiled records for games, such as Minesweeper and cup stacking. We also looked at the evolution of growth rates in the long-term evolution experiment involving {\it E. coli} \cite{lenski1994dynamics}. In total, we analyze 171 datasets reflecting 5278 records, and only include record progressions with at least ten new records. (Those interested in accessing the data should contact the authors). Overall, this diversity of records is intended to capture a more comprehensive set of the possible forms that record progressions can take, possibly shaped by distinct constraints, selective pressures, and interactions with cultural and technological innovation.

\subsection{Summary statistics}

In order to describe this space and reveal differences in record setting processes, we develop a set of summary statistics. Clearly, there are many candidate summary statistics that could be developed. Some, like the mean time between record breaking events or mean percent improvement on the previous record, depend on the overall scale of time over which records are kept or the scale over which the underlying values vary. For records that are approaching an asymptotic limiting value, the rate of improvement will be decelerating. However, even though the rate at which the record changes over time is slowing down, the rate at which new records are set might be continuing apace or even accelerating, as seems to be the case for some speed skating events. The distribution in time of record breaking events is insensitive to the range over which the underlying values vary (it is invariant under monotonic transformations). Therefore, summary statistics of this distribution --- such as whether the rate of events is accelerating, whether events are clustered (``burstiness''), and the correlation between subsequent waiting times --- can be more readily compared between record breaking progressions across domains.

\begin{figure}
\centering
\includegraphics[width=\textwidth]{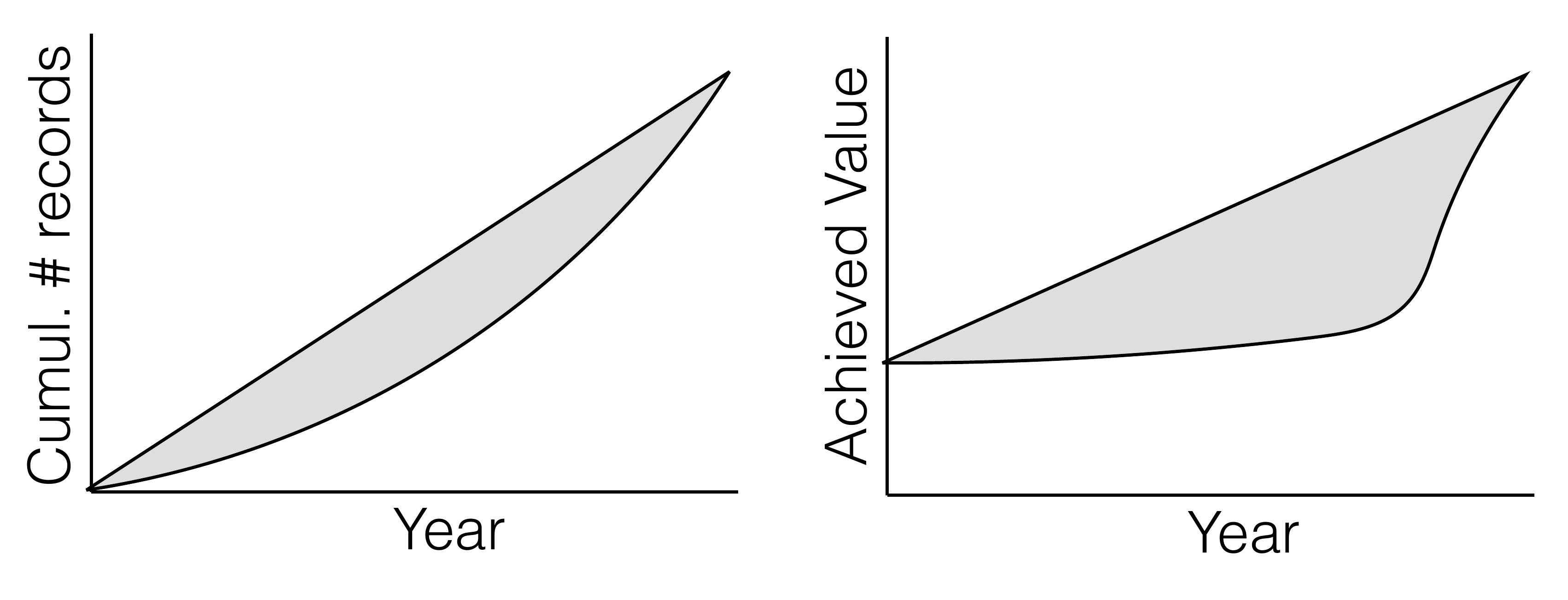}
\caption{\label{fig:schematic} \textbf{Schematic illustration of the ``Record-setting acceleration'' and ``Value acceleration'''€ measures.}  Left) For a given record progression, we plot the cumulative \# of records (scaled between 0 and 1) achieved over time.  Record-setting acceleration is the area between a line indicating a linear growth of record counts and the actual growth of record counts.  It is positive when record setting is speeding up over time, and negative when it is slowing down over time.  Right) For a given record progression, we plot the maximal value achieved (scaled to have a maximum of 1) over time.   Value acceleration is the area between a line indicating linear growth of the record value and the actual record-breaking value. }
\end{figure}

For each given record progression, we collected interarrival times between records. We then computed the following set of statistics:
\begin{enumerate}
\item {\bf Improvement rate:} the percentage that the extremal value increased per new record $(r_m/r_1)^{1/m}-1$, where $r_1$ is the first record established, $r_m$ is the last, and $m$ is the number of records. This was clipped at 25\% for visualization purposes. 
\item {\bf The ``coefficient of variation'' of interarrival times:} the standard deviation of the interarrival times divided by the mean $\mathrm{Var}(w)^{1/2}/E(w)$, where $w_i=t_{i+1}-t_i$ is the length of time that the $i$th record stood without being broken. This is a standard measure of the ``burstiness'' of a process in, for example, queuing theory and  neuroscience.
\item {\bf The ``acceleration'' of record breaking events:}  To measure this, we plotted the cumulative number of record breaking events against the date at which they occurred.  We then measured the signed area between linear growth and this curve, normalized by the total variation of the two axes (see Figure \ref{fig:schematic}).  This is positive when records arrive at an increasingly fast rate and negative when they arrive at a decreasing rate.
\item {\bf The acceleration of the record-breaking value:} As the previous metric does for the rate of record-breaking events, this metric measures the acceleration of the rate of change of the record itself. Again, we compare to a linear change over time and take the signed difference in area between the observed curve and the linear curve. See Figure \ref{fig:schematic}.
\end{enumerate}
Note that to compute these measures in a uniform way, all record progressions were rescaled to grow over time.  For example, instead of the time to run a 100 meter dash, we use the inverse, proportional to average speed over 100 meters.

\begin{figure}
\centering
\includegraphics[width=\textwidth]{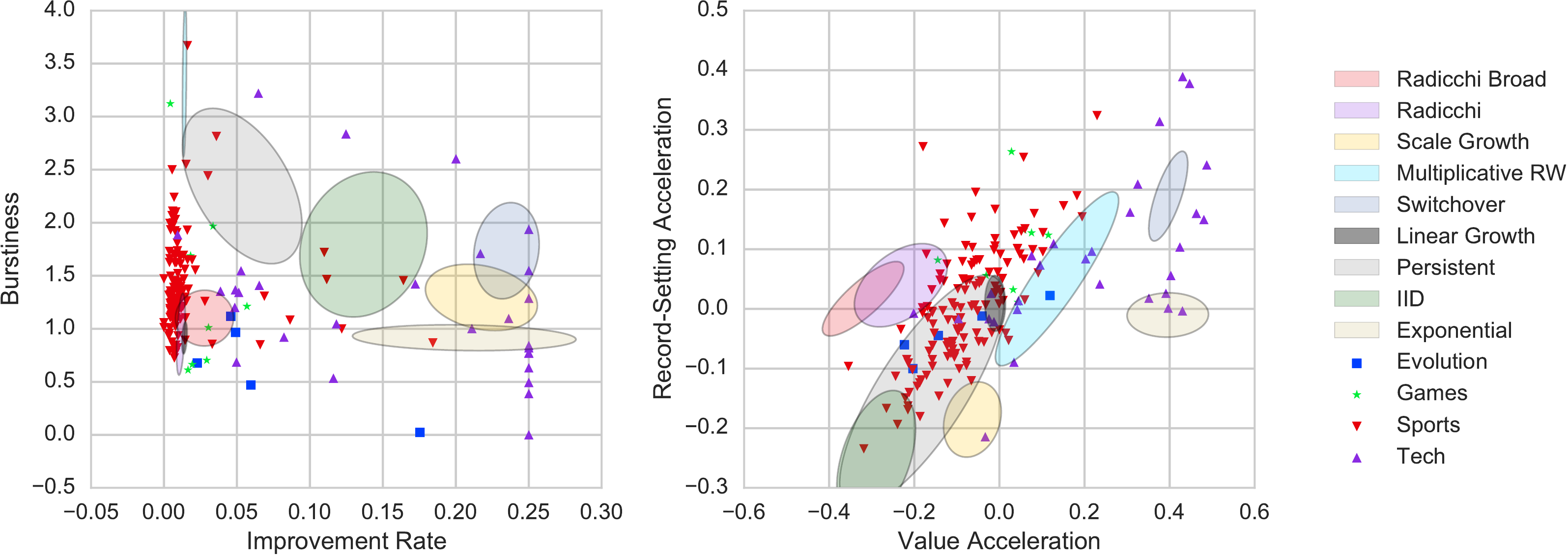}
\caption{\label{fig:scatters} \textbf{Phase diagram of various kinds of record progressions.} Points reflect individual record progressions, colored by category as sports, biology, games (video games, etc.) and technological progressions. Shaded regions cover various generative models.}
\end{figure}

Deploying these metrics, we are able to extract a few high-level observations regarding the phenomenological range of record progressions from various domains. We find that most sports records improve very slowly (see improvement rate in Figure \ref{fig:scatters}a). Games (video games, cup stacking) improve faster than sports, followed by examples of experimental biological evolution, and finally, technology grows fastest. In terms of burstiness, sports populate a wide range, as does technology. The burstiest technological examples are largest pumpkins and tallest buildings. It is interesting to note that in both of these examples, have relatively low total improvement rates, and underwent a qualitative regime shift from incremental, infrequent progress to large, frequent progress. The experimental evolution examples have much lower levels of burstiness. This is because improvements tend to come at a regular pace owing to the very large population sizes and a fixed large number of generations between sampling performance.

In terms of record acceleration, we again find that the largest pumpkins and tallest buildings are the most striking. For sports, the lowest acceleration values occur for the sports where our initial records are close to the inferred asymptotic value such as horse racing and the 100-meter dash for men. Some technological advances, such as the cost of gene sequencing and supercomputer performance, experience steady exponential growth. This behavior is exhibited in terms of high value acceleration (due to exponential growth) and vanishing record-setting acceleration (due to consistent year-to-year improvement).

In order to further classify the phenomenological range, we generated ranges of values for each of the metrics using a few generative models. Each of these models generates a sequence of values over time, of which we only keep the ones that are larger than all previous values of the sequence.

We use a variety of generative models to capture both classical theory and the range of phenomena we observe in empirical data (see Section~\ref{sec:background} for references and further discussion of the properties of these models):
\begin{enumerate}
\item Perhaps the simplest generative model is a process of repeated, independent draws from a fixed distribution: $x_t \sim X$, $t=1,\ldots,T$. This type of process is often referred to as \textit{IID} or \textit{discrete white noise}. We use an exponential distribution. In terms of our metrics, this generative model does not capture almost any of the observed progressions. (Labeled ``IID'' in Figure~\ref{fig:scatters}).

\item A model that corresponds to persistent change in the range of performance is a process of linear increase with IID exponentially distributed noise: $x_t \sim X + v t$. Depending on the relative scale of the stochastic noise and the deterministic trend, this process can exhibit a variety of behaviors (Labeled ``Linear Growth'' in Figure~\ref{fig:scatters}).

\item In many technological domains, the underlying progress is exponential. Therefore, the exponential of the previous process, giving an exponential increase with multiplicative noise, is a useful generative model. (Labeled as '€œExponential'€ in Figure~\ref{fig:scatters}).

\item Many cumulative processes are characterized by successive values building on their immediate predecessors. A multiplicative random walk, biased toward increasing, captures this short-memoried cumulative process with long-term progress: $x_{t+1} \sim x_{t} \exp(N(\mu,\sigma))$. We label the range of metrics we observed from this model as ``Multiplicative RW'' in Figure~\ref{fig:scatters}.

\item One explanation suggested for the lack of deceleration in the rate of setting new records that is predicted from the IID model is that while the underlying performance distribution does not change, the rate at which draws from this distribution are made is growing exponentially, due to, for example, an exponentially growing population of competitors: $x_t = \max_{i=1,\ldots,N_t} x_{t,i}$, $x_{t,i}\sim X$. We do not show the range of metrics observed for this model in Figure~\ref{fig:scatters}.

\item A related model is that the scale of the distribution is increasing monotonically with time: $x_t \sim (c + v t)X$.  (Labeled as '€œScale Growth'€ in Figure~\ref{fig:scatters}).

\item A phenomenological model proposed by Radicchi, based on performance in Olympic athletic events, models the approach to an asymptotic limiting value as a multiplicative random walk, biased downwards, subtracted from a fixed offset \cite{radicchi_universality_2012}. We have two ranges corresponding to this model in Figure \ref{fig:scatters}: ``Radicchi'' labels the range observed for multiple realizations of the process using parameters given in Ref.~\cite{radicchi_universality_2012} for the 400-meter men's running race; ``Raddicchi Broad'' labels a range from a range of randomly sampled parameters.

\item A few of the record progressions we collected, such as the tallest-building or largest-pumpkin records, are characterized by a distinct technological breakthrough or regime shift. To model these generativity, we use a process that starts as repeated IID draws from an exponential distribution. At some point in time, the scale of the distribution from which draws are made starts growing exponentially. The range for this model is labeled as ``Switchover'' in Figure~\ref{fig:scatters}.

\item In many fields, the progression of records is dominated by repeated record setting by a few individuals. For example, of the 55 record-breaking performances in the women's pole vault, 48 were made by only 4 individuals with 10 or more record-breaking performances each. Similar patterns occur for the women's hammer throw and the men's 200 meter butterfly swimming race. Since this individual-dominated effect is not captured by any of the previous generative models, we also created an individual-level model: each individual in a population of fixed size is modeled as a distribution of typical performances, with some individual mean and exponential noise. We start with a population of $N$ individuals, and at each time step, we draw one sample from the distribution associated with each individual. We use the maximum of those as the winning value for that time step. (This is similar to the growing-population model above, as studied in Yang \cite{yang_distribution_1975}, except that here the population is allowed to be heterogeneous, instead of every individual having the same distribution.) A fixed number of individuals are discarded and redrawn at each time step by sampling their mean performance from an exponential distribution. The range corresponding to this model is labeled as ``Persistent'' in Figure~\ref{fig:scatters}. However, these assumptions may be violated in real world datasets. A more detailed consideration of the effect of individual performance is given in the next section (see Section \ref{sec:marathon})
\end{enumerate}

\section{Course-Graining: The Marathon}\label{sec:marathon}

The marathon, with hundreds of thousands of competitors around the world annually, is a rich source of data for testing the assumptions of record-progression models, and in particular, the consistently high performance of individual athletes. At the most fundamental level, records in any competition are set by a population of competitors, and thus, record-progression models make assumptions about that population of competitors to derive their predictions. With finishing times of hundreds of elite athletes, historical data from major marathons in particular are particularly well suited to test model assumptions. They also reveal the extreme reduction in information that occurs when entire competitions are reduced to the statistics of their winners, and when those statistics are again reduced into a record progression. 

To better test the mechanistic assumptions of record-progression models, we examined the top-100 male and top-100 female finishers of five of the annual world major marathons (New York City, Boston, Chicago, London and Berlin; data from the Tokyo marathon were not accessible online; see Methods). Selecting the top 100 competitors in particular enabled us to focus on individuals with the potential to break a course record or a world record. The marathon data has the additional benefit that the number of participants in each race is much larger compared to other competitions (e.g., 100m dash, which are usually conducted in batches of eight), so the elite pool of individuals who could be in contention for breaking records is larger than in many other competitions. 

\begin{figure}
    \centering
        \includegraphics[width=\textwidth]{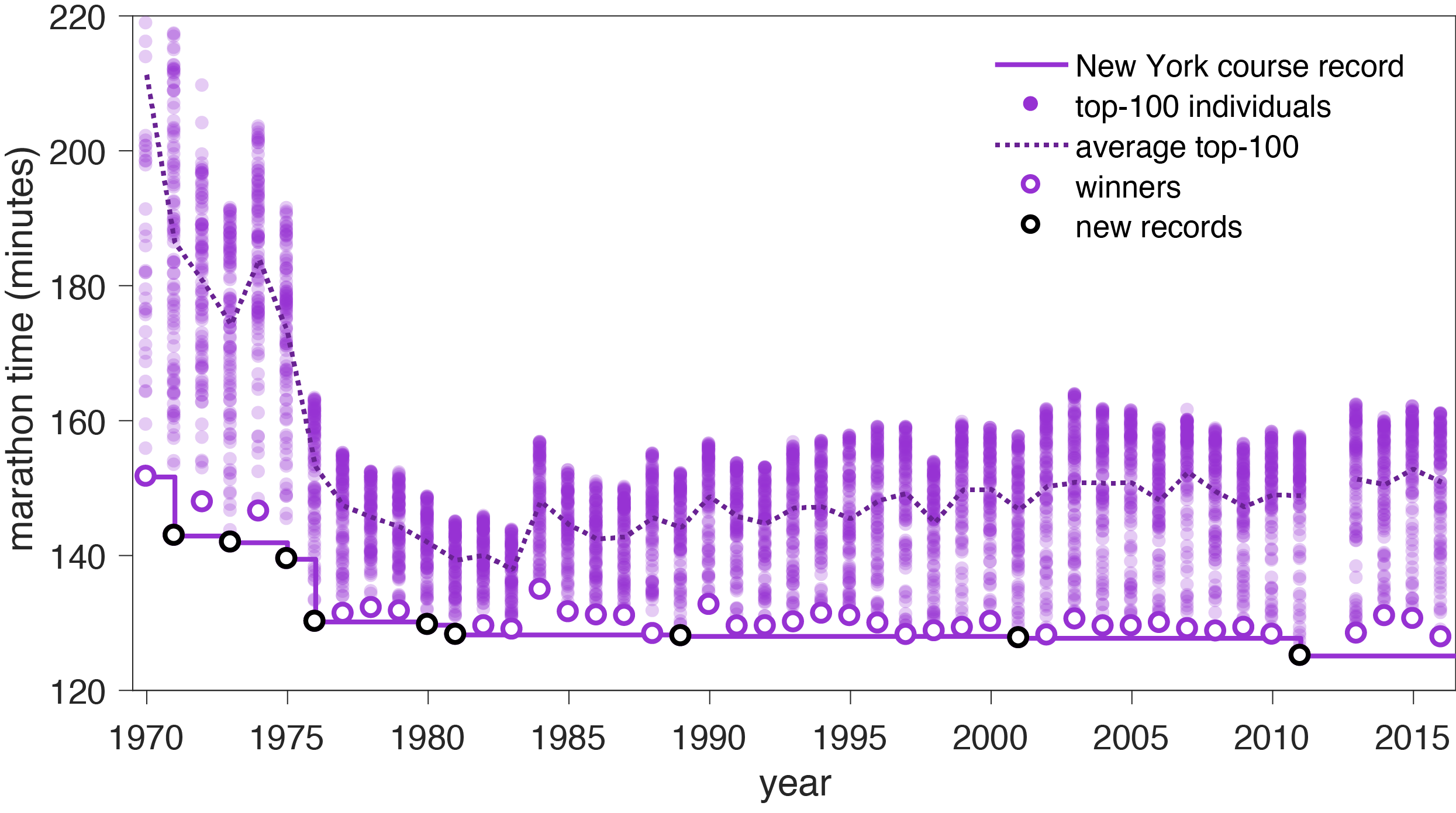}
    \caption{{\bf Information lost by looking at the record progression time series and ignoring the evolution of the underlying distribution---NYC Marathon}. Each point represents a top 100 finisher in a given year, the dotted line connects the means of each years distribution of finishers. The bottom line depicts the running course record for the New York City Marathon and a black circle represents when that record was set. While the record progression time series (the black dots) appear roughly unchanged after 1975, with only slight improvements, the progression of distributions above this line provides a more complete story. Showing that while the records are not evolving the field of competition is. For example, the mean and variance begin increasing after 1983 suggesting an ever increasing spread between top 100 and podium level competitors, something that is lost in the record progression alone but may be meaningful in understanding the stagnation in record progression.}\label{fig:nyc-information-loss}
\end{figure}

Marathon data reveal multiple mechanisms, many of which are hidden by the record progression summary statistics. For example, after falling 15\% in its first 5 years, the course record for the New York City Marathon fell only an additional 5\% decrease over the next 40 years (Figure~\ref{fig:nyc-information-loss}). Yet the annual distributions of top-100 finishers reveal far more detailed information about the finishing times that have produced (or not produced) new course records. In particular, there is phase of rapidly improving annual performance and a collapse in the range of competitive ability among top New York runners in the early 1980s, while years thereafter show a much broader set of elite finishing times. To understand record-breaking marathon running without understanding these patterns may defy the predictions of simple models. 

\subsection{Finishing Distribution Analysis}\label{subsec:finishing_dist_anal}
The traditional analysis framework for record breaking assumes a fixed number of draws from a stationary distribution, yet direct estimation of this stationary distribution is difficult without data. In our marathon data, we examine the connections between this random draw perspective and one more contingent on the structure of the underlying data. 

In marathons, the actual number of entrants in each race varies from year to year, violating the fixed-draw assumption, though it is unclear \emph{a priori} if this change is significant. At the same time, the vast majority of marathon runners are neither making serious attempts to win nor capable of winning\footnote{One could think to use the entire distribution of the ``elite'' marathoners, but these data can be difficult to acquire and often elite runners will end their race early if they are having a poor race}. However, if it is (close to) true that the ``serious'' distribution is stationary and there are an equal number of draws, then the distribution of the top-100 finish times should also be stationary.

\begin{figure}
\centering
  \includegraphics[width=0.47\linewidth]{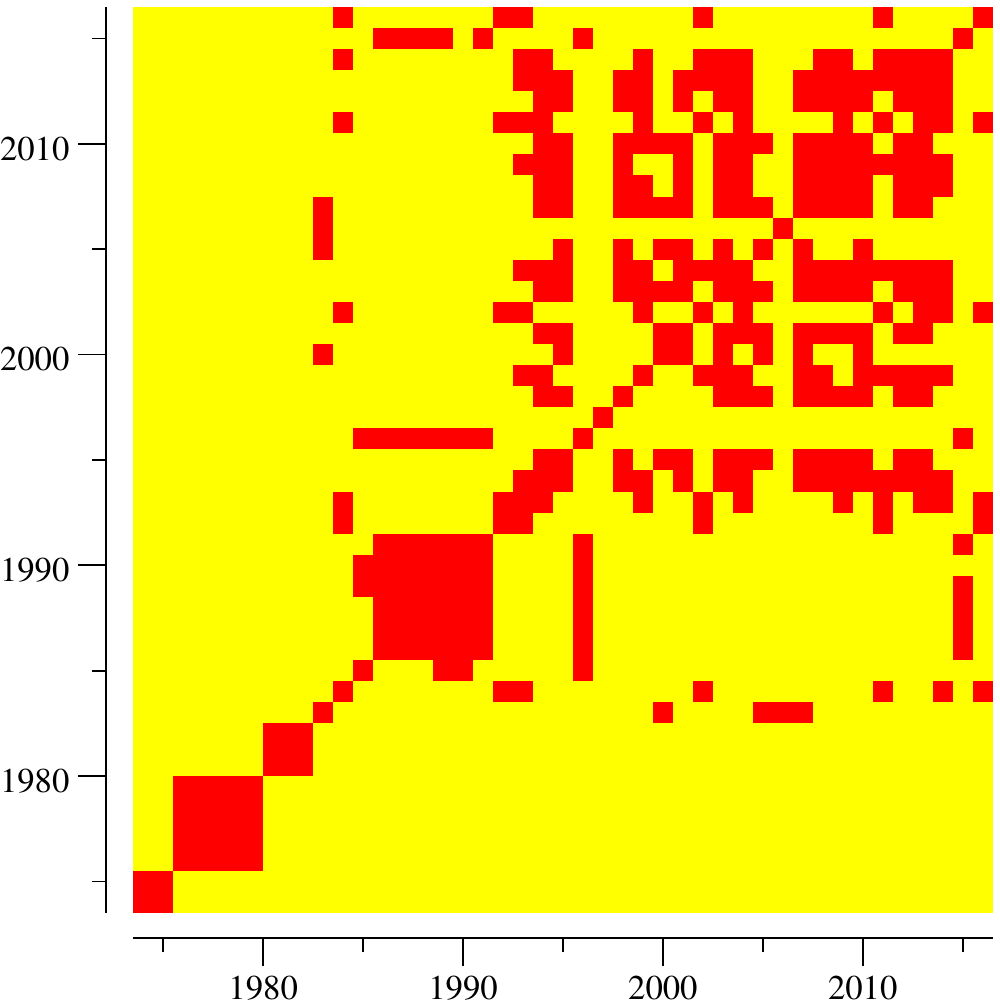}    \includegraphics[width=0.47\linewidth]{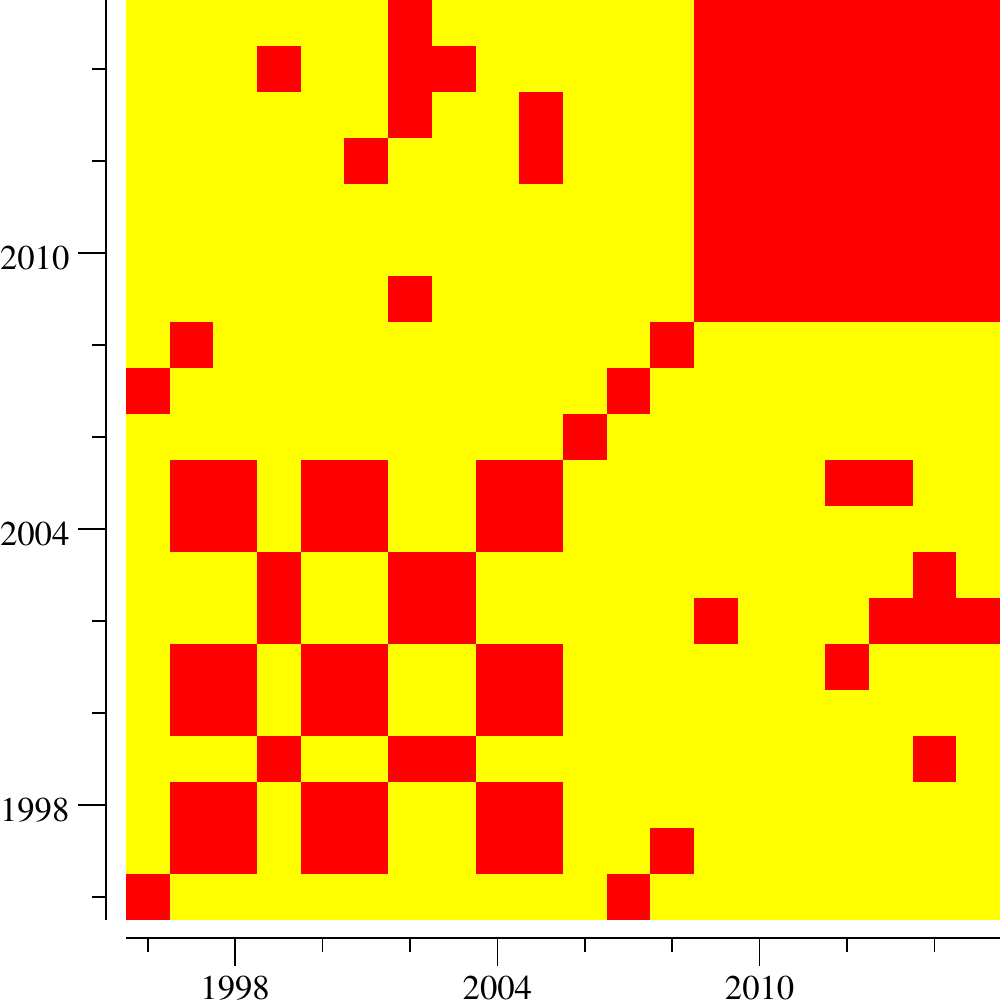}  \\[6pt]
 \includegraphics[width=0.47\linewidth]{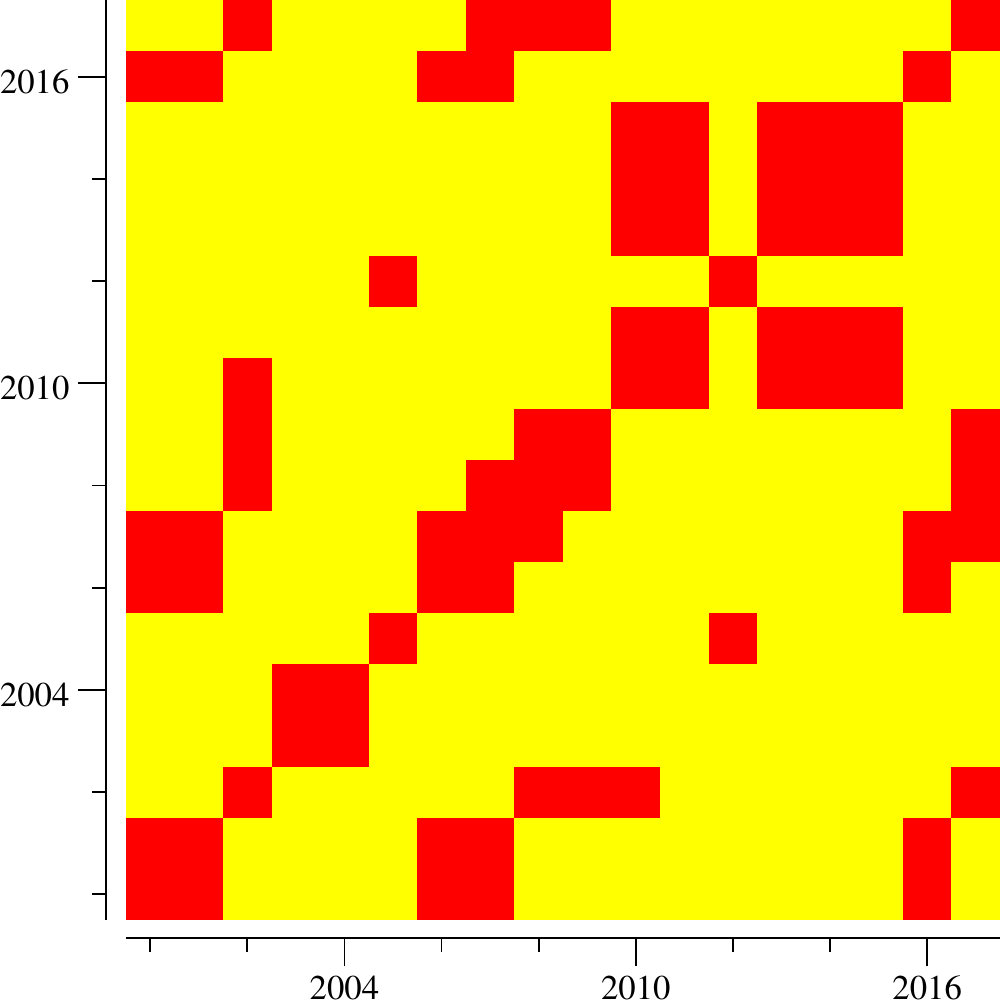}    \includegraphics[width=0.47\linewidth]{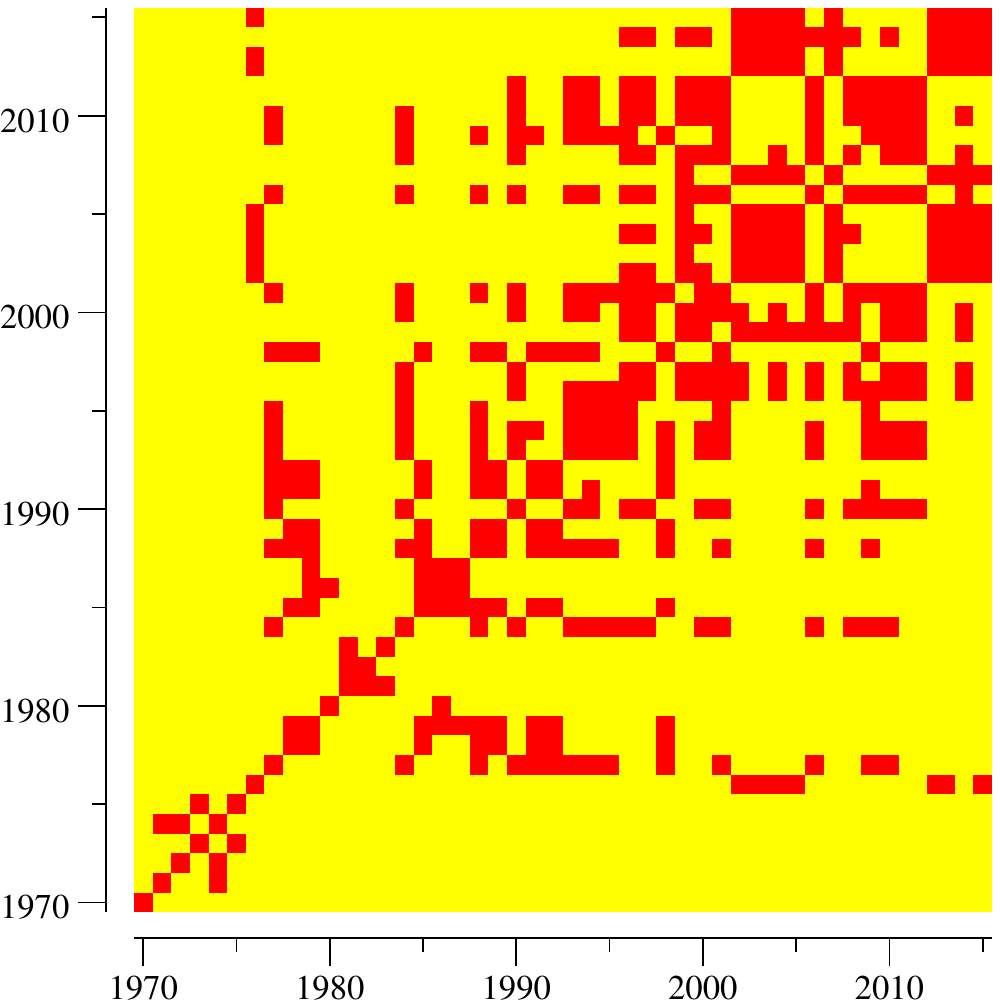}  \\[6pt]
\caption{{\bf Kolmogorov-Smirnov plots} from left to right and top to bottom: for Berlin, Chicago, Boston, and New York. Yellow refers to $p$-values lower than 0.01, and thus a rejection of the hypothesis that the top-100 finish times of the corresponding year-pairs are generated from the same distribution. The KS-plot for Berlin (top-left) suggests several regime shifts. An early period (pre-1984), where finishing times are significantly different from later years, and mostly different from one another, a second period (1984-1993) where the distribution was stationary but distinct from earlier and later years, and a third period (1998 onward), where the distribution seems to have stabilized with a few outliers. The New York City marathon (lower left) is less clearly delineated, but seems to show a similar trend with early years being distinct from all other years, and later years being similar to one another with some outliers. The Chicago (top-right) and Boston (lower-left) marathons are less clear in their interpretations, but both have large regions of non-rejected year-pairs, especially post 2010.}\label{fig:heatmap}
\end{figure}

We test the stationarity hypothesis by comparing the year-to-year top-100 finishing times for the each of the five marathons. In particular, we compute the two-sided (two-sample) Kolmogorov-Smirnov test~\cite{birnbaum_numerical_1952} for each pair of years in each marathon, asking whether there is sufficient evidence to reject the hypothesis that they are generated from the same distribution ($p < 0.01$; Figure~\ref{fig:heatmap}).

The earliest years of each marathon are significantly different from years that follow. This suggests that current performance in the marathon is indeed well-modeled by a stationary stochastic process. Note that this analysis \emph{does not} take into account exogenous factors that are known to affect marathon performance, such as weather \cite{ely2007impact,suping1992study}. Correcting for these factors would likely make marathon outcomes look less statistically distinct from one another, and confirms the obvious: exogenous factors have significant impact on the likelihood of setting future marathon records. It will be interesting to see if recent efforts to run a sub-two-hour marathon will impact this apparent stationarity, and provide evidence for a statistically-significant shift in marathon performance.

\begin{figure}
    \centering
        \includegraphics[width=0.7\textwidth]{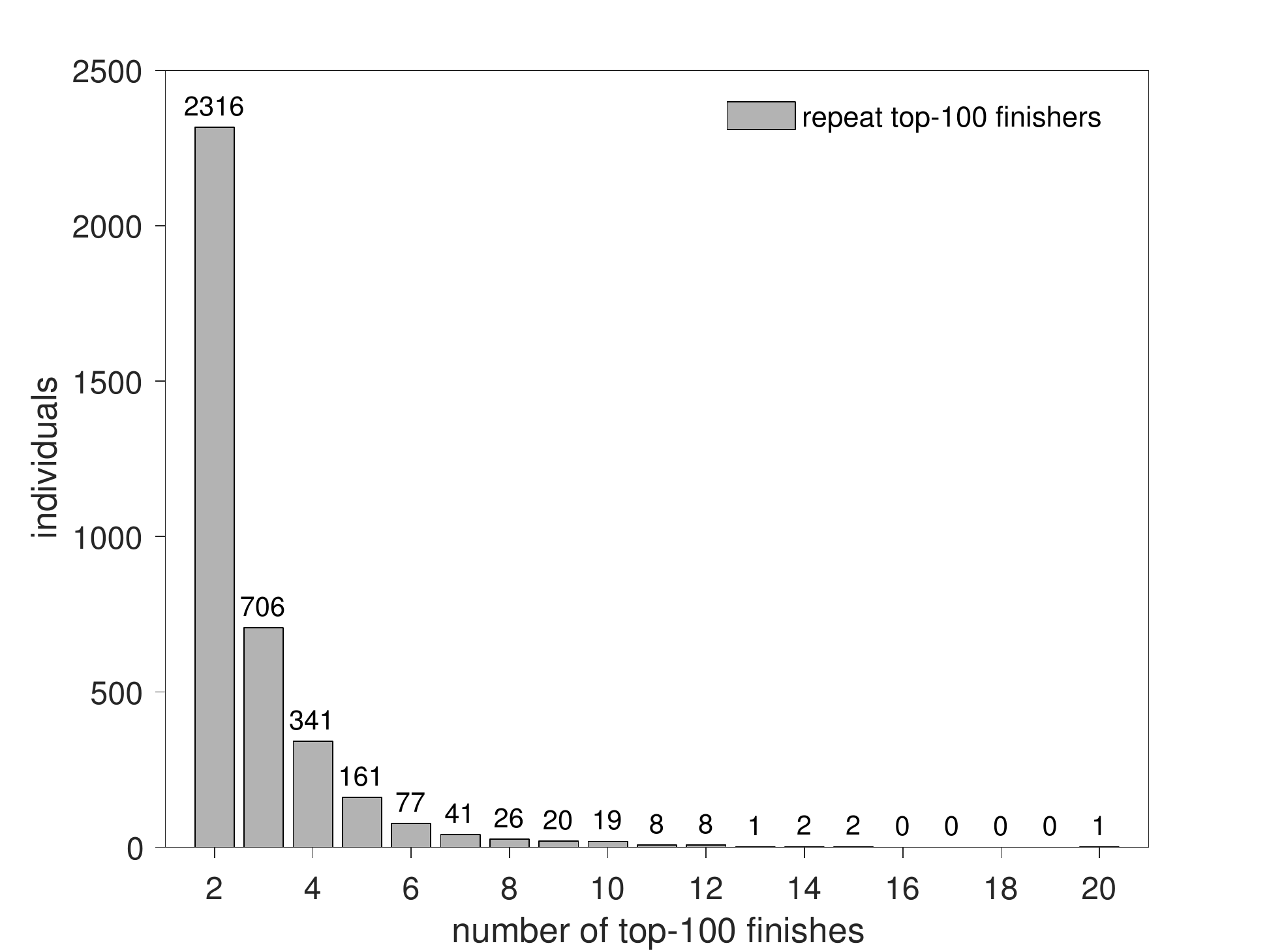}
    \caption{{\bf People are not anonymous particles.} This is a histogram over the number of top 100 finishes per person over all five datasets. It shows that there is a large number of runners, who participate in multiple races, and hence, it shows the non-independent nature of the major marathons over the year. Independence is one of the underlying assumptions of many models found in literature. The data suggests that other methods are required.
}\label{fig:nyc_marathon_men_repeat}
\end{figure}

\subsection{These particles have names}
For many of the record-progression datasets analyzed in this paper, individual record holders appear multiple times (e.g., Sec.~\ref{subsec:finishing_dist_anal} and Fig.~\ref{fig:nyc_marathon_men_repeat}). Due to this fact alone, the perspective of competitors as anonymous particles, randomly drawn from some set distribution, is missing a critical component of actual competition that could be affecting record progressions: these particles have names. 

Across major marathons, many individuals repeatedly finish in the top 100. We used our person ID codes (see section~\ref{sec:marathon-methods}) to track individuals across time and races. In contrast to the Kentucky Derby, in which 100\% of horses are newcomers to the event each year, we found that 13.5\% of top-100 women and 14.5\% of top-100 men were also in the top 100 in the previous year. In fact, simply counting the number of top-100 finishes for each runner reveals that individuals are often phenomenally successful (Figure~\ref{fig:nyc_marathon_men_repeat}): among men, Mebrahtom "Meb" Keflezighi had the highest number of top-100 finishes in our dataset, finishing within the top-100 men in 15 marathons (and winning both the New York City and Boston marathons). Even more astounding is the performance of Gillian Horovitz, who finished within the top-100 women in 20 races in our dataset. (Horovitz's top-100 placements in major races should be even higher, but not captured in our data due to availability.)

This non-independence of race-to-race participation by elite runners indicates that the statistical assumptions of traditional record modeling are not met. It is not true that in every event an entirely independent set of statistics is drawn. Instead, there are correlations between the year-to-year top 100 finishers. It is interesting to see that, despite these correlations, the overall statistics of top marathon finishers seem to be more similar than different. Future work is needed to understand the mechanisms that generate these different regimes in marathon performance. What kind of distributions over individual human trajectories lead to very different year-on-year results as in the early days of the marathon? What distributions lead to the apparent current situation, where individual humans improve and decline in their performance over time, and yet the overall statistics remain the same? If we can understand the underlying statistics and mechanisms by individual trajectories combine, we should be able to improve upon the mechanism-free approaches that dominate in record prediction.

\section{Discussion}

In this work, we consider a broad range of record-setting phenomena in different domains, such as sports, games, biological evolution, and technological development.  Progressions in these domains are characterized by different statistical patterns and underlying processes. Record progressions in some domains agree very well with previous models and assumptions, while others populate a new space, and require novel models and assumptions. 

Here we found that cross-cutting metrics allowed us to organize record progressions in terms of their dynamical properties. To characterize dynamical properties we employed measures of rates of improvement, ``burstiness'' of record-breaking time series, and the acceleration of the record breaking and the acceleration of record values. We found that records in the same general domain tend to cluster together. For example, sports and games exhibit the slowest rate of improvement while technology records improve at a much faster rate and show a strong acceleration. 

In order to connect these macroscopic statistics to detailed processes, we examined marathon results, where we found that studying record-setting data alone can hide many of the structural properties of the underlying the process. The marathon study also illustrates how some of the standard statistical assumptions used in record progression models may be routinely violated in real-world datasets.

The full distribution of elite marathon runners---those that come close to, but did not necessarily break a record---provides a more complete picture of the record setting process. For example, our analysis showed that simply considering record breaking events, and hence reducing a rich dataset to extremal summaries, removes a substantial amount of information, which could be critical in detecting the underlying mechanisms leading to record progressions as well as our ability to predict who will break a record, and when. Our study of marathon data also illustrates that some of the standard statistical assumptions underlying record progression models such as IID draws from some stochastic process may be inappropriate or commonly violated by real-world datasets.

Individual trajectories are clearly important to the science of record-breaking. Under some conditions these seem to combine to match previous simplifying assumptions about record-breaking, while in other cases seem to combine to yield substantially different dynamics. Further study is needed to understand when the contributions from individual trajectories define an overall pattern inconsistent with previous models.

\newpage
\section*{Methods and Supplemental Information}

\subsection{Waiting time} \label{sec:waitingtime}

For a stochastic process, the record will improve after certain events. The time interval that passes between record setting events is known as the waiting time. If the generative process for outcomes is independent draws from a stationary distribution from an exponentially growing population, then the distribution of waiting times between record-setting events is the exponential distribution, with the base of the exponent being the population growth rate (as discussed in \ref{sec:background}). 

We test the hypothesis of exponentially distributed waiting times for the collected record progressions in the dataset. For each dataset, we compute a list of waiting times, and we fit the maximum-likelihood (ML) best exponential to the data (rate parameter set to the inverse of the mean waiting time) \footnote{The best-fit exponential is based only on observed waiting times, and does not consider how long the current record has been held. For many of the collected datasets, we are not sure that record keeping has continued to the present, and thus how long this ``unobserved'' waiting time has actually been measured.}. We then use an Anderson--Darling test \cite{anderson2011anderson} to measure if the ML exponential is likely to have generated the observed waiting time distribution. Specifically, we compute the $p$-value for each data set being generated under the exponential based on the Anderson--Darling statistic. 

Across the records, about $35\%$ were found to be significantly not-exponential $(p < 0.01)$, and about $20\%$ of them are \emph{very} significantly non-exponential $(p < 10^{-6})$. For the remaining two-thirds of the samples, we could not reject the null hypothesis of an exponential distribution ($p > 0.01$), with nearly half having $p$-values greater than $0.10$. The shape of the distribution thus matches that expected from IID draws from an exponentially growing population (see \ref{sec:background}). However, the rate parameters we find are all quite small, corresponding to growth rates of at most $0.8\%$, suggesting that while population effects are present, there are significant additional phenomena occurring, even when we cannot rule out the exponential distribution. Interestingly, all of the evolutionary data sets have exponentially distributed waiting times, except for one; the difference between that one and the others is that it covers five times fewer generations (2,000 instead of 10,000), at five times higher resolution (sampled every 100 generations instead of every 500). 

\subsection{Marathon analysis methods}\label{sec:marathon-methods}

We collected data on marathon finish times for the top 100 men and women finishers in five major marathons: Boston, Berlin, Chicago, London, and New York. We collected data using a combination of methods (entering data by hand and by scraping online datasets via python using selenium as a webdriver and for HTML parsing). We collected data from multiple online repositories: Boston, Chicago, London race data were all scraped from  \href{http://www.marathonguide.com/index.cfm}{http://www.marathonguide.com/index.cfm}, Berlin race data was collected by hand from \href{http://www.bmw-berlin-marathon.com/en/facts-and-figures/results-archive.html}{http://www.bmw-berlin-marathon.com/en/facts-and-figures/results-archive.html}, and supplemental race data was collected from Wikipedia as needed. 


We wanted to identify individual racers across different years or different races, but databases differed in the format and reporting for names. Some databases, like \href{http://www.bmw-berlin-marathon.com/en/facts-and-figures/results-archive.html}{http://www.bmw-berlin-marathon.com/en/facts-and-figures/results-archive.html}, only reported each individual's first initial and their last name, while most other databases reported full first names and last names. In order to link these individuals across datasets, despite differences in formatting, we coded each person's name as first initial and last name. We then conducted extensive data testing to disambiguate individuals which now were coded with the same abbreviated name code, but which may be different individuals. We flagged records if (1) a coded name was recorded multiple times as finishing in the top 100 finishers of the same race, (2) a coded name had an unusually long career (especially those individuals with unusually long apparent career breaks between top-100 finishes). An example are runners with running careers longer than 10 years but with less or equal to three appearances. For these runners, we manually checked additional data, such as age and nationality for flagging. (3) Online available data from professional runners suggests that it is very unlikely that a runner participates in more than two races a year. We only found one such runner in our data, and could verify from news sources, that he had run all six major races in one year.

\subsection{Additional course-graining: more marathon data}

\begin{figure}
    \centering
	\includegraphics[width=\textwidth]{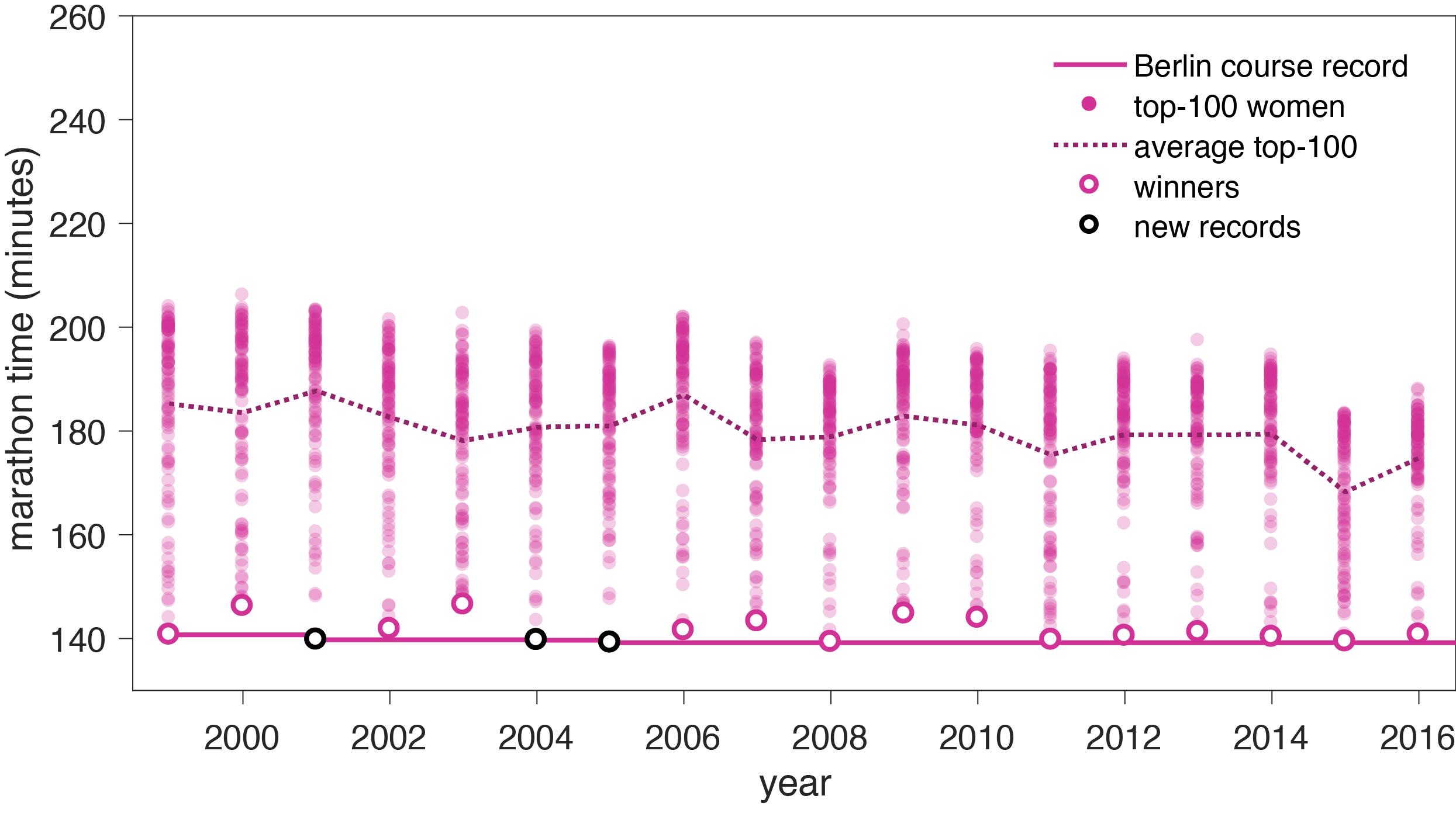}
        \includegraphics[width=\textwidth]{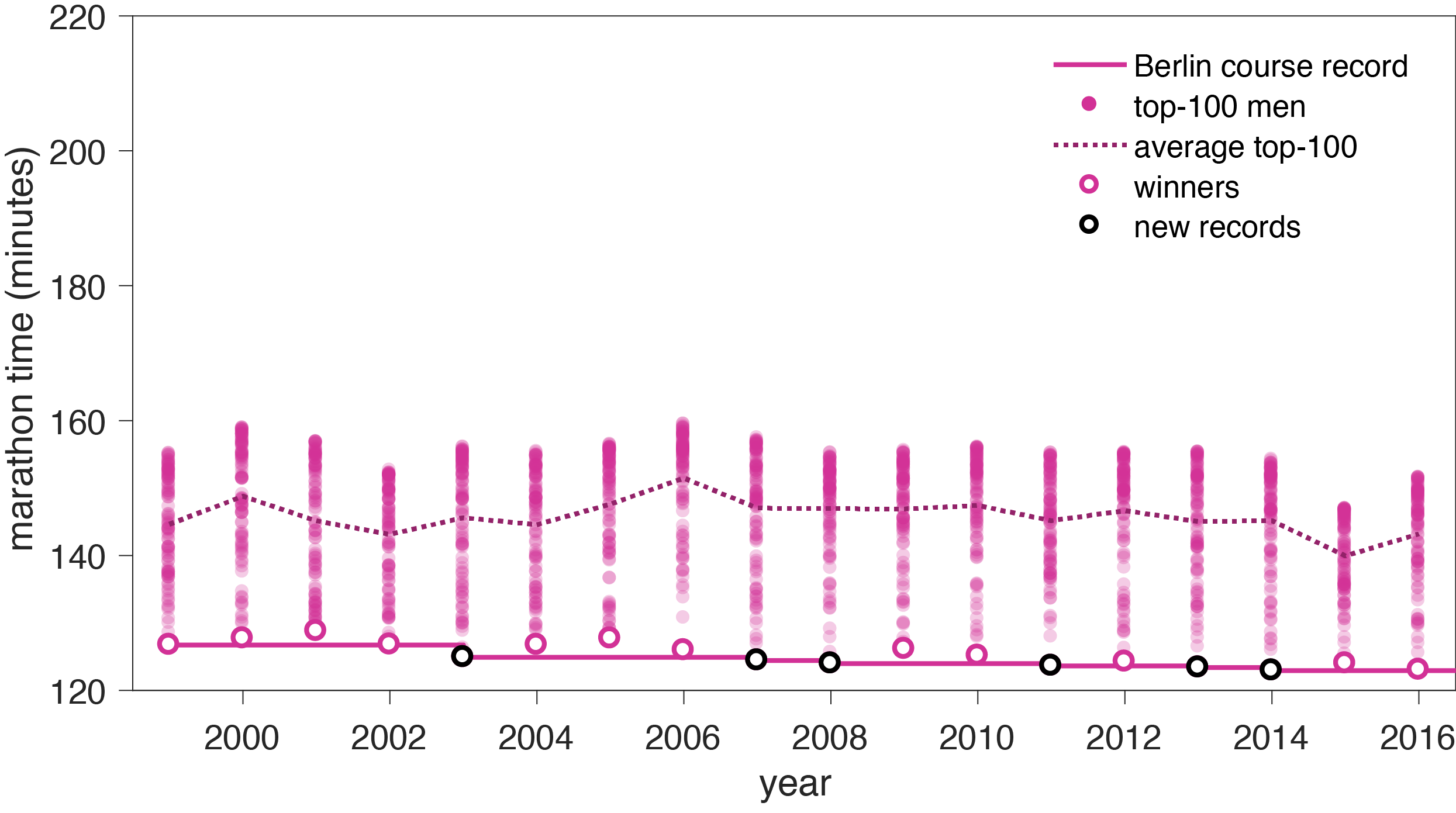}
    \caption{{\bf Berlin Marathon} data for women (top) and men (bottom).}
\end{figure}

\begin{figure}
    \centering
	\includegraphics[width=\textwidth]{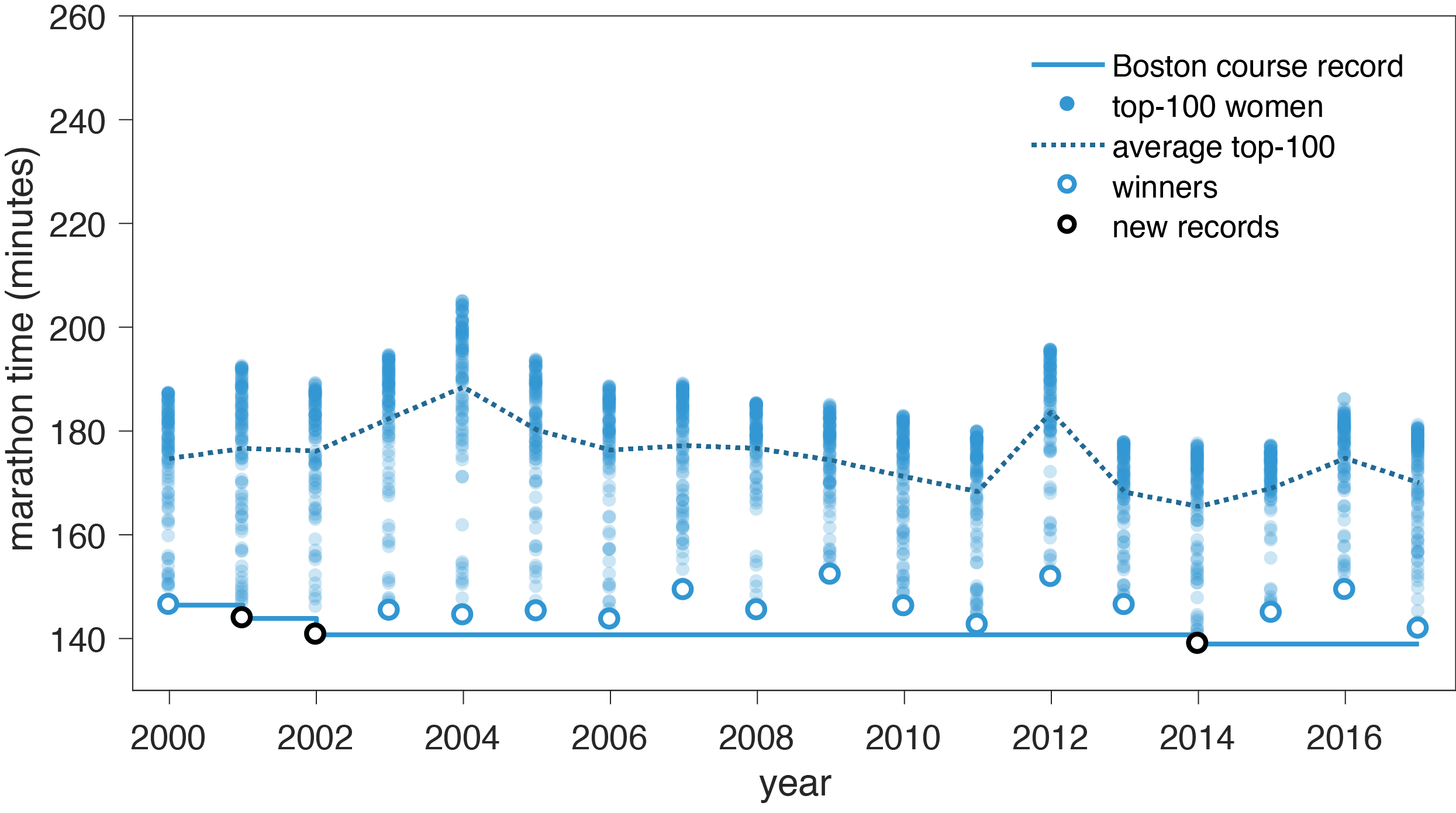}
        \includegraphics[width=\textwidth]{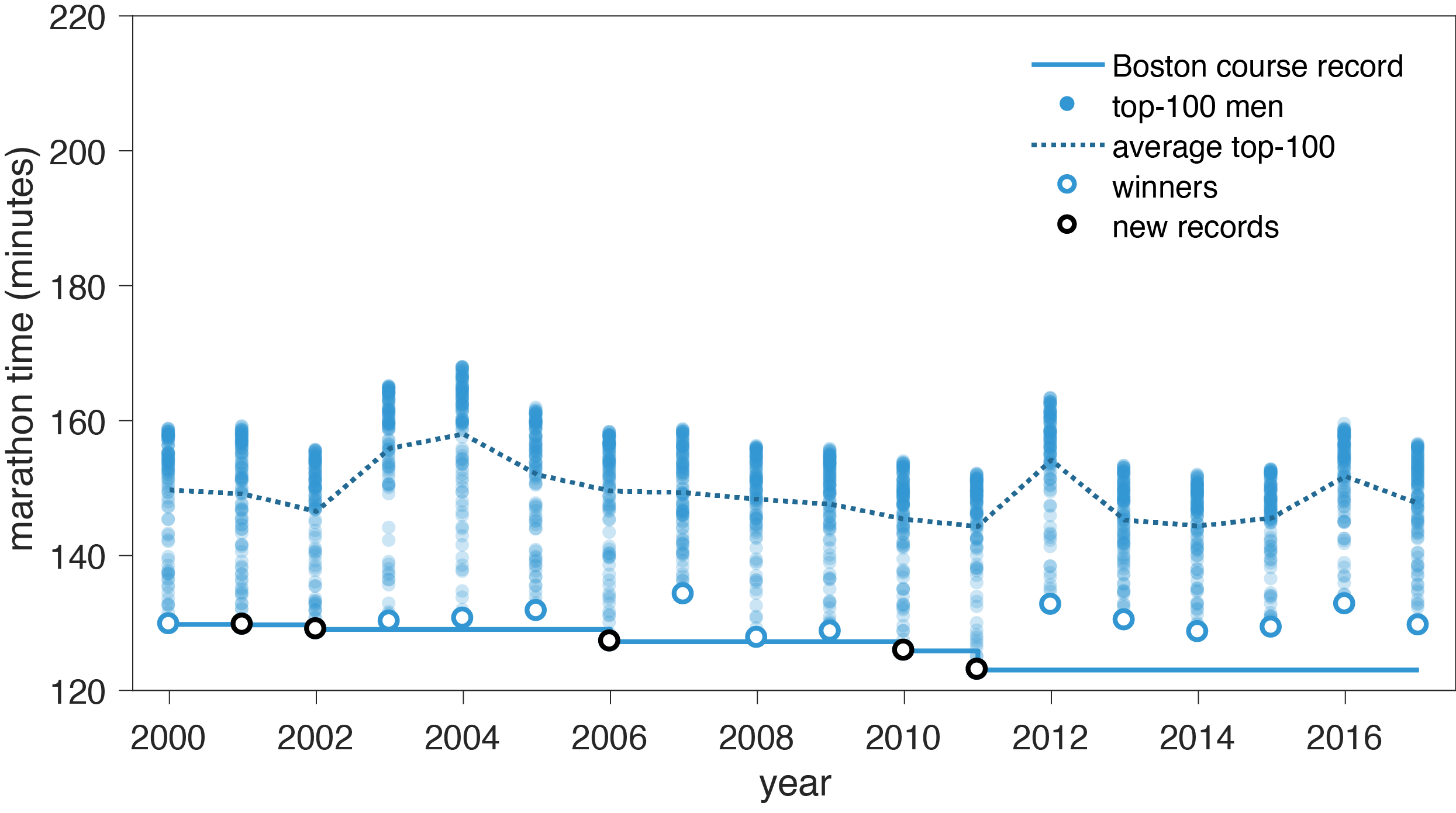}
    \caption{{\bf Boston Marathon} data for women (top) and men (bottom).}
\end{figure}

\begin{figure}
    \centering
	\includegraphics[width=\textwidth]{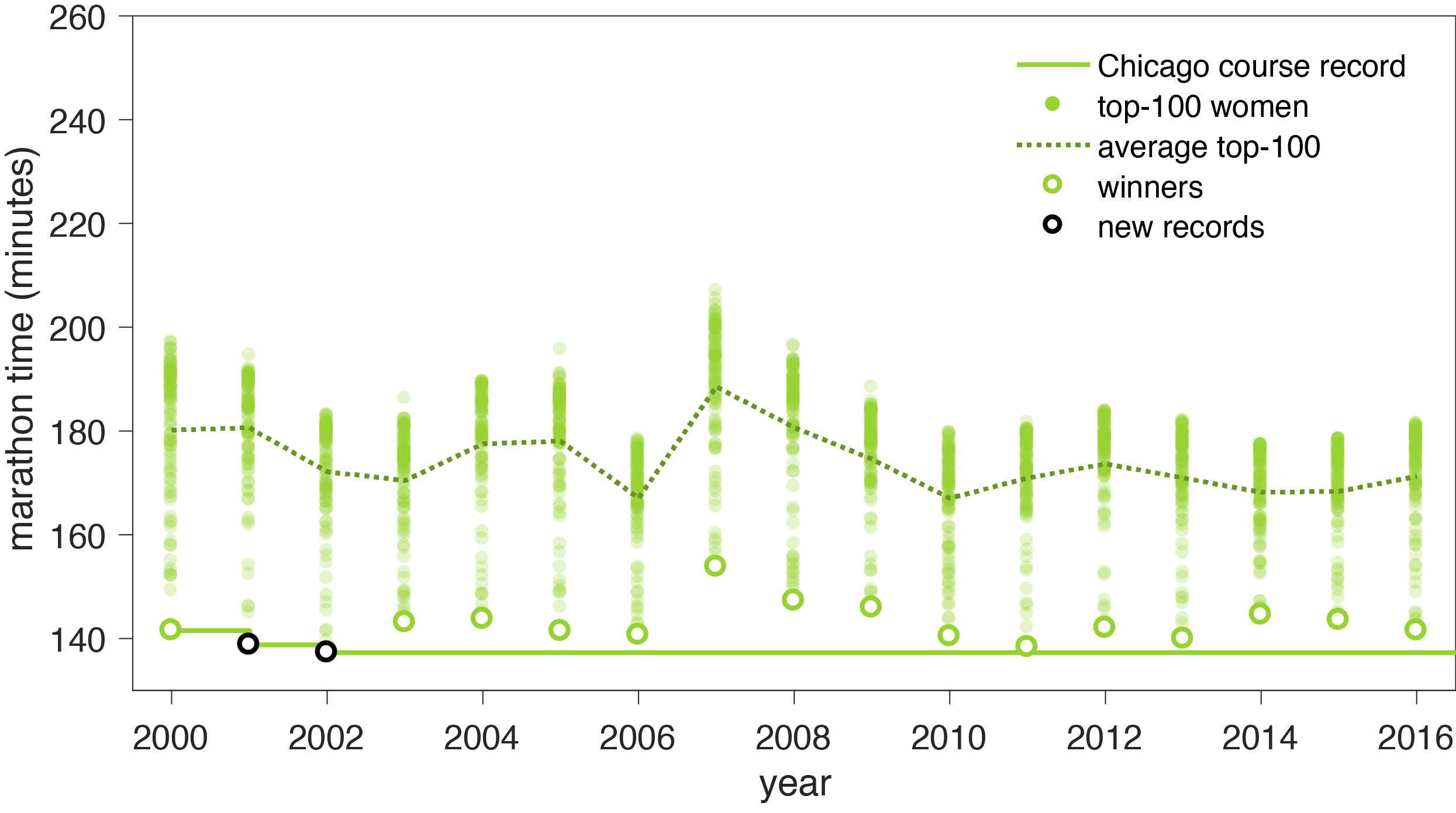}
        \includegraphics[width=\textwidth]{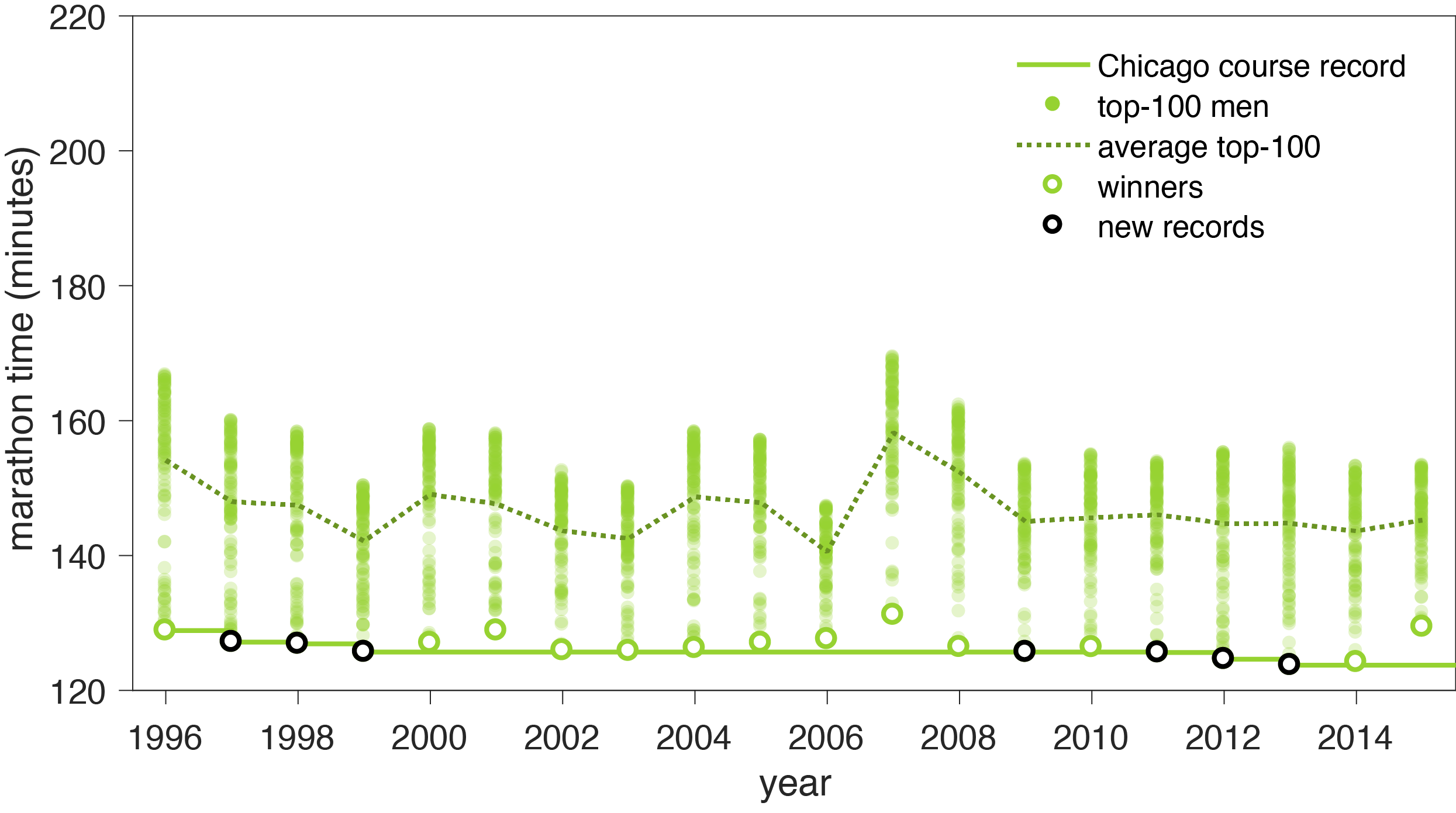}
    \caption{{\bf Chicago Marathon} data for women (top) and men (bottom).}
\end{figure}

\begin{figure}
    \centering
	\includegraphics[width=\textwidth]{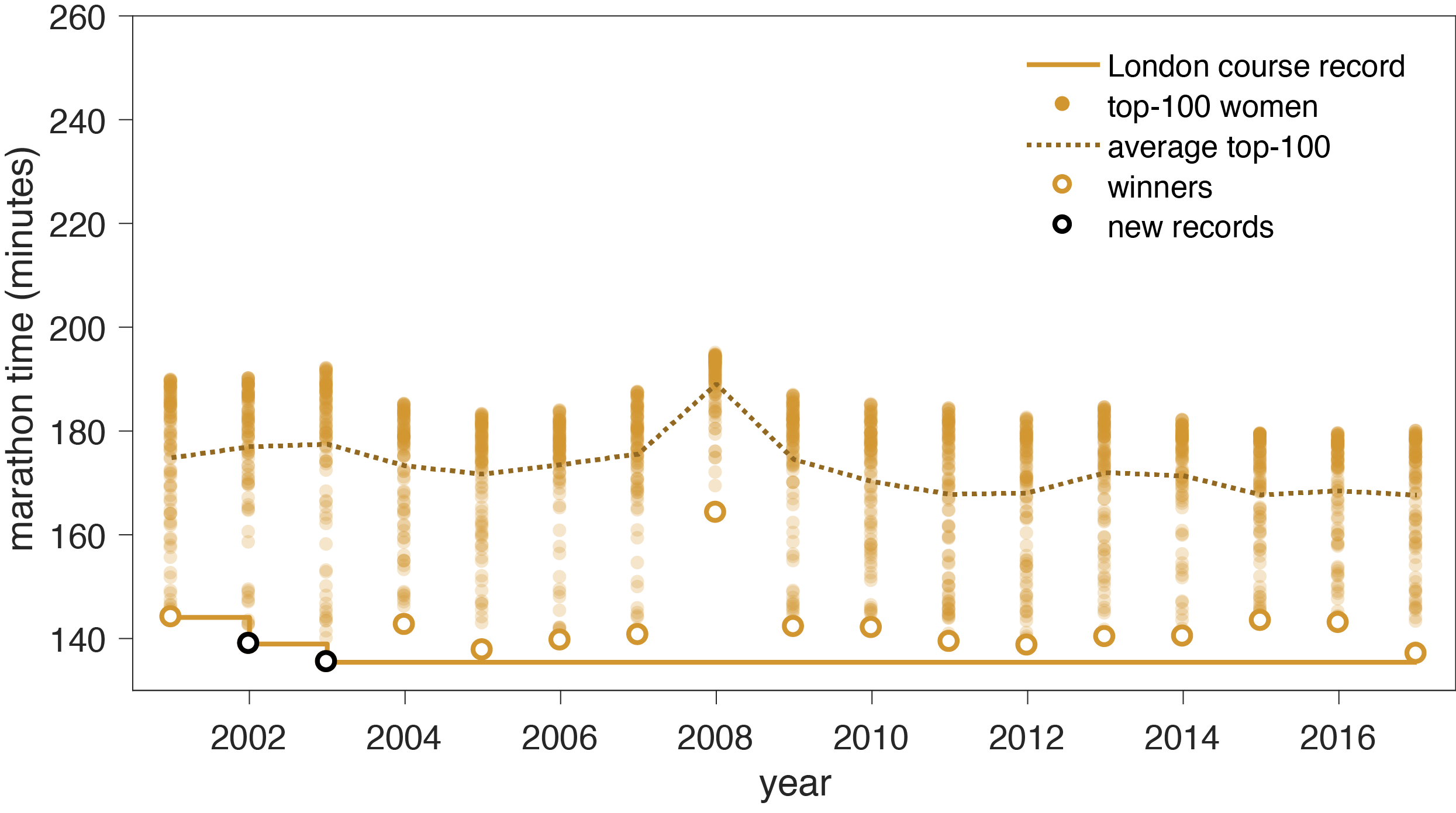}
        \includegraphics[width=\textwidth]{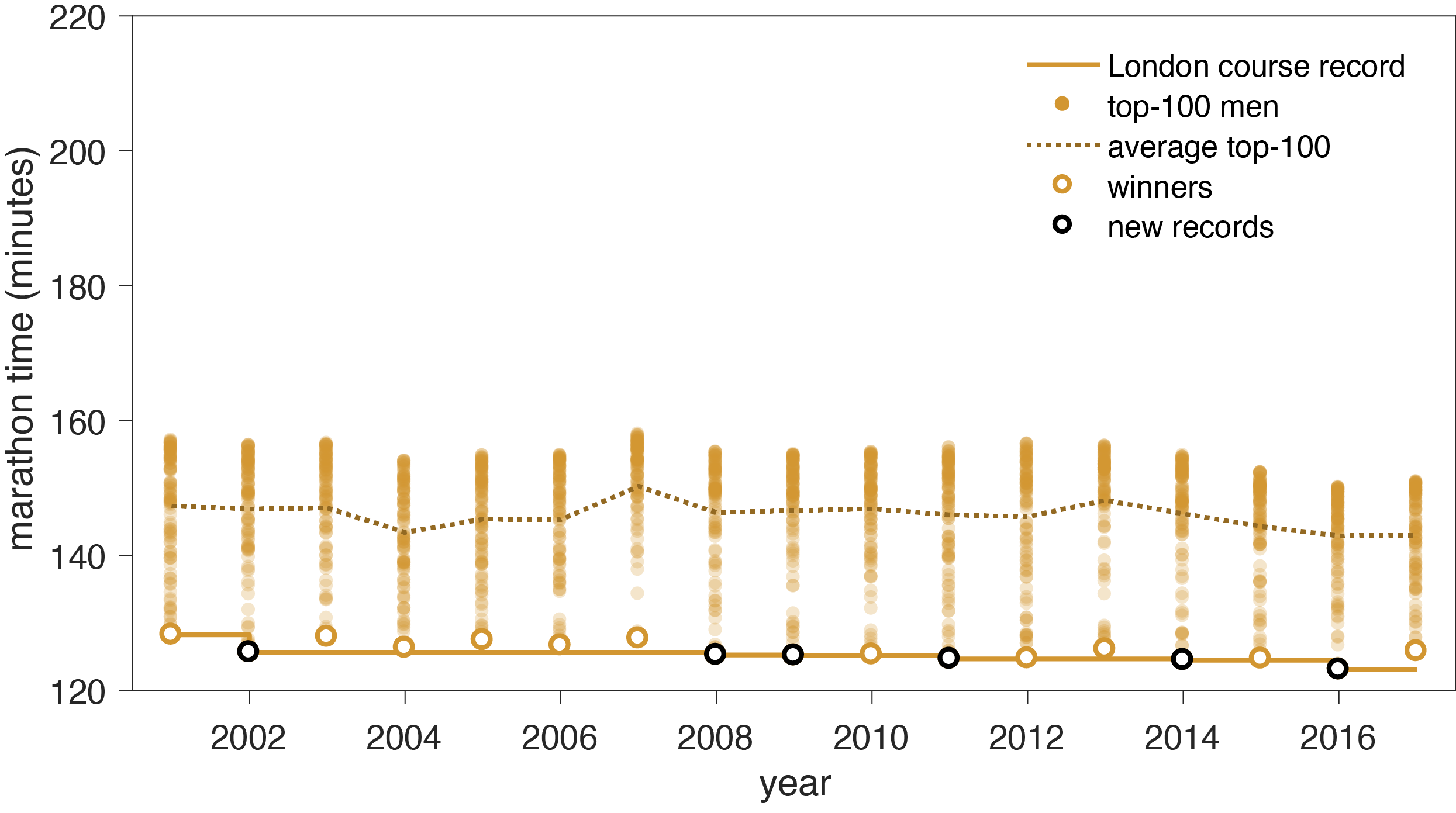}
    \caption{{\bf London Marathon} data for women (top) and men (bottom).}
\end{figure}

\begin{figure}
    \centering
	\includegraphics[width=\textwidth]{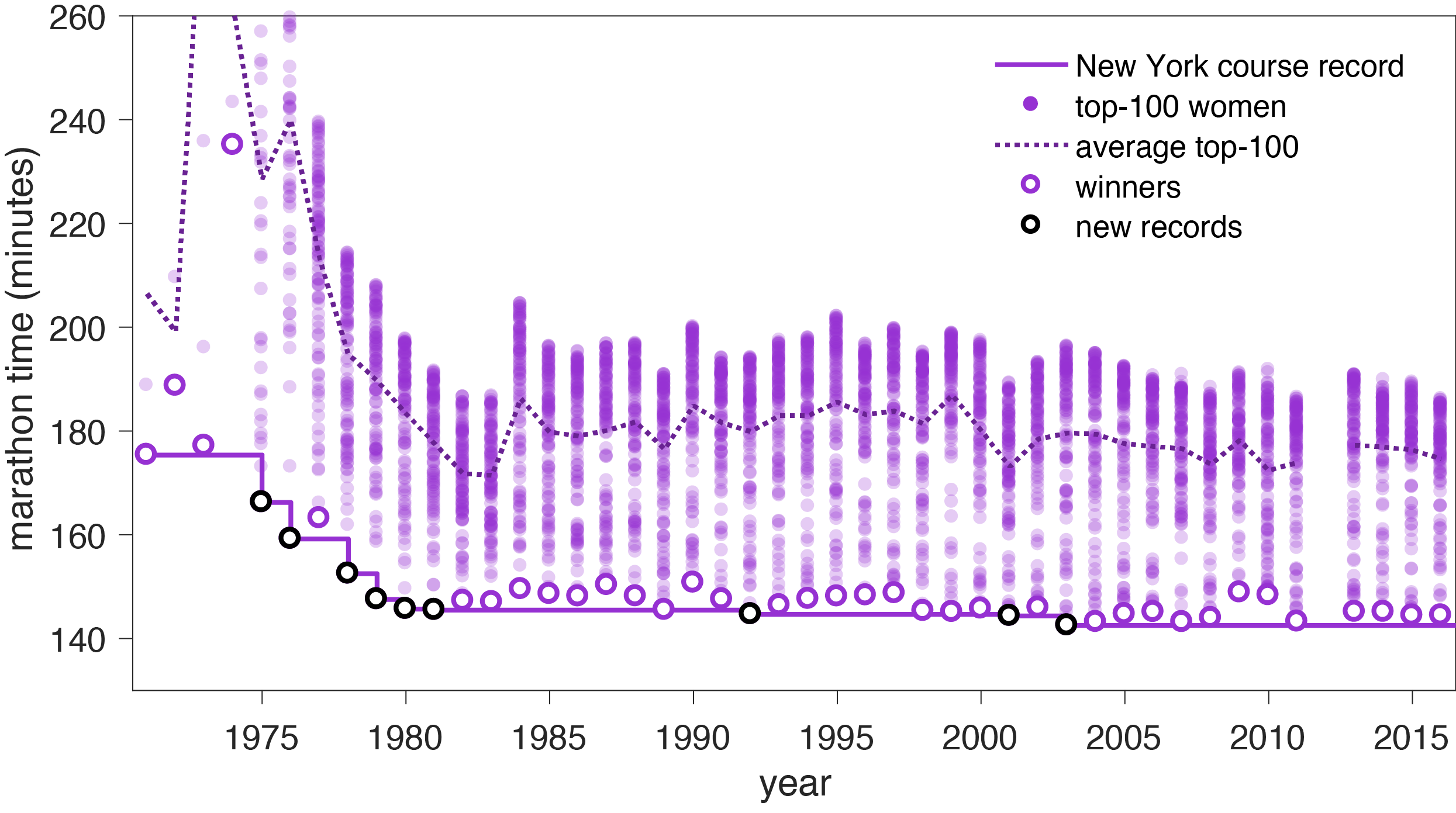}
        \includegraphics[width=\textwidth]{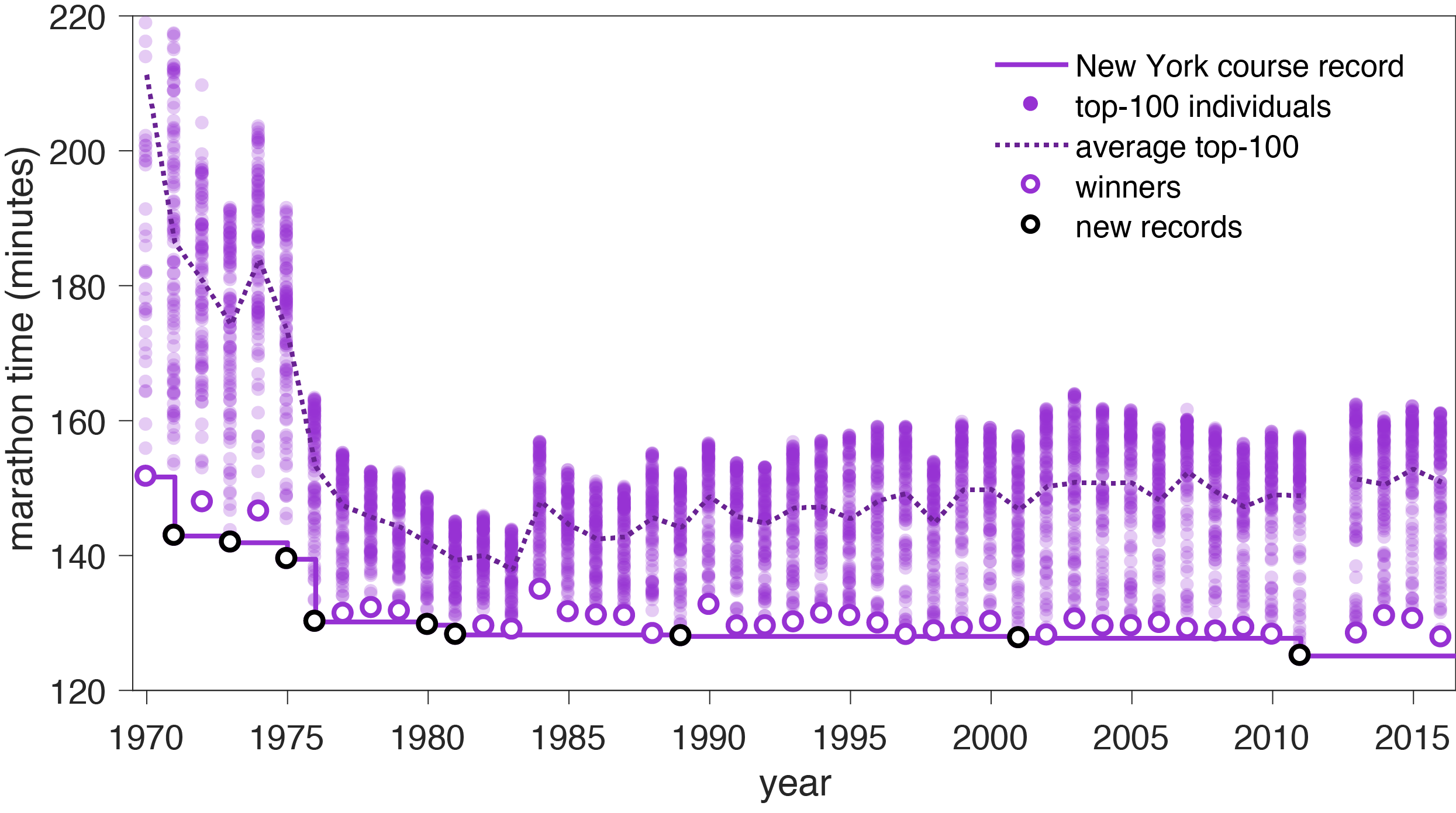}
    \caption{{\bf New York Marathon} data for women (top) and men (bottom).}
\end{figure}

\clearpage
\appendix

\section{72 Hours of Science}\label{statement}

During the 72 hours between 1 pm on May 8 and 1 pm on May 11, 2017, postdocs from the Santa Fe Institute (SFI) will be conducting an experiment. This is both a description of that experiment's particulars and an open letter declaring our commitment to those particulars. It will also be included as a supplement to any eventual publication.

For most projects that we work on as postdocs, science moves slowly. Research takes months or years to evolve from early-stage ideas to coherent and crisp findings, with additional months of writing before submitting work for publication. During the 72 Hours of Science, we will actively challenge the notion that this is the only timescale for science by going from idea to submitted arXiv post in only 72 hours. Our goal is also to engage nearly the entire postdoc community at SFI with specialities including anthropology, pure mathematics, urban development, cultural evolution, physics, computer science, ecology, political science, and evolutionary biology. Thus a second basic aspect of this informal experiment is to test the limits of science not only in terms of pace but also multidisciplinarity.  We think it'll be interesting, challenging, and fun. The agreed upon rules are as follows:

\subsection*{Rule 1: Submit to the arXiv within 72 hours}
The minute the clock starts, at 1:00 pm MDT on May 8, 2017, we will begin to choose a project to work on. Before the clock stops, at 1:00 pm MDT on May 11, 2017, we will submit our written manuscript to the arXiv, along with a copy of this description which registers the informal 72h(S) experiment. During the intervening three days, we will decide on a project, develop the project, divide tasks among the participating postdocs, and write up our findings. It should be noted that the arXiv post will not be a description or analysis of the informal experiment, but of the generated scientific project itself. This document, which you're reading now, is intended to catalog the process as a supplement to the scientific paper generated from the process.

\subsection*{Rule 2: Fresh ideas only}
Choosing a project will be key to the success of 72h(S), so in order to select from the most diverse set of ideas, any participant can propose a project idea to the group. However, in the spirit of a good challenge, only projects or ideas that are new to everyone are allowed---they cannot be discussed beforehand, nor can they be substantially developed by the presenter. In other words, as a group, we're starting from scratch with a fresh idea and then developing it together. Since we did not want to rule out ideas that were mentioned, or created, during an off-hand discussion or comment during a seminar we have the formal criterion that three-quarters of the group should not have heard the problem statement before, and we allowed each person to informally ``bounce'' the idea off one person not at SFI.

\subsection*{Rule 3: We're in this together}
It's a collaboration, not a competition. Over the course of three days, each participant will occupy multiple roles, from reading to writing, coding to reviewing, cooking to coffee-making.  So, regardless of whose fresh idea is chosen, authorship will be Santa Fe Institute Postdocs for the arXiv submission, with an alphabetical list of authors as the first reference with a statement of uniformly equal credit.

\vspace{0.2in}
\begin{center}
	Agreed upon by SFI Postdocs on May 8, 2017, 12:55 PM MDT.
\end{center}

\section*{Acknowledgments}
For useful discussion of related ideas we thank David Krakauer and Ryan O'Donnell. We are grateful for generous financial support from the Omidyar family and the Santa Fe Institute. We thank the Santa Fe Institute for support in executing the 72 Hours of Science event. We especially thank Bruce Bertram, Rebecca Bradshaw, Jennifer Dunne, David Krakauer, Jenna Marshall, Nathan Metheny, John Miller, Hilary Skolnik, and Scott Wagner.

\bibliographystyle{unsrt}
\bibliography{refs}

\begin{thebibliography}{10}

\bibitem{noubary_survival_2004}
Farzad Noubary and Reza Noubary.
\newblock On survival times of sport records.
\newblock {\em Journal of Computational and Applied Mathematics},
  169(1):227--234, August 2004.

\bibitem{radicchi_universality_2012}
Filippo Radicchi.
\newblock Universality, limits and predictability of gold-medal performances at
  the {Olympic} {Games}.
\newblock {\em PLOS ONE}, 7(7):e40335, July 2012.

\bibitem{redner2001guide}
Sidney Redner.
\newblock {\em A guide to first-passage processes}.
\newblock Cambridge University Press, 2001.

\bibitem{chandler_distribution_1952}
K.~N. Chandler.
\newblock The distribution and frequency of record values.
\newblock {\em Journal of the Royal Statistical Society. Series B
  (Methodological)}, 14(2):220--228, 1952.

\bibitem{aitken_statistical_2004}
Tony Aitken.
\newblock Statistical analysis of top performers in sport with emphasis on the
  relevance of outliers.
\newblock {\em Sports Engineering}, 7(2):75--88, 2004.

\bibitem{berthelot_athlete_2010}
Geoffroy Berthelot, Muriel Tafflet, Nour~El Helou, St\'ephane Len, Sylvie
  Escolano, Marion Guillaume, Hala Nassif, Julien Tola\"ini, Val\'erie
  Thibault, Fran\c{c}ois~Denis Desgorces, Olivier Hermine, and
  Jean-Fran\c{c}ois Toussaint.
\newblock Athlete atypicity on the edge of human achievement: {Performances}
  stagnate after the last peak, in 1988.
\newblock {\em PLOS ONE}, 5(1):e8800, January 2010.

\bibitem{berthelot_citius_2008}
Geoffroy Berthelot, Val\'{e}rie Thibault, Muriel Tafflet, Sylvie Escolano,
  Nour~El Helou, Xavier Jouven, Olivier Hermine, and Jean-Fran\c{c}ois
  Toussaint.
\newblock The {Citius} end: world records progression announces the completion
  of a brief ultra-physiological quest.
\newblock {\em PLOS ONE}, 3(2):e1552, February 2008.

\bibitem{charalambides_distribution_2007}
Charalambos~A. Charalambides.
\newblock Distribution of record statistics in a geometrically increasing
  population.
\newblock {\em Journal of Statistical Planning and Inference},
  137(7):2214--2225, July 2007.

\bibitem{einmahl_records_2008}
John H.~J. Einmahl and Jan~R. Magnus.
\newblock Records in athletics through extreme-value theory.
\newblock {\em Journal of the American Statistical Association},
  103(484):1382--1391, 2008.

\bibitem{gembris_sports_2002}
Daniel Gembris, John~G. Taylor, and Dieter Suter.
\newblock Sports statistics: {Trends} and random fluctuations in athletics.
\newblock {\em Nature}, 417(6888):506--506, May 2002.

\bibitem{gembris_evolution_2007}
Daniel Gembris, John~G. Taylor, and Dieter Suter.
\newblock Evolution of athletic records: statistical effects versus real
  improvement.
\newblock {\em Journal of Applied Statistics}, 34(5):529--545, 2007.

\bibitem{glick_breaking_1978}
Ned Glick.
\newblock Breaking records and breaking boards.
\newblock {\em The American Mathematical Monthly}, 85(1):2--26, 1978.

\bibitem{holmes1969note}
Paul~T Holmes and William~E Strawderman.
\newblock A note on the waiting times between record observations.
\newblock {\em Journal of Applied Probability}, 6(3):711--714, 1969.

\bibitem{kelley2006predicting}
Daniel~J Kelley, Jonas~R Mureika, and Jeffrey~A Phillips.
\newblock Predicting baseball home run records using exponential frequency
  distributions.
\newblock {\em arXiv preprint physics/0608228}, 2006.

\bibitem{krug_records_2007}
Joachim Krug.
\newblock Records in a changing world.
\newblock {\em Journal of Statistical Mechanics: Theory and Experiment},
  2007(07):P07001, 2007.

\bibitem{majumdar_universal_2008}
Satya~N. Majumdar and Robert~M. Ziff.
\newblock Universal record statistics of random walks and {L\'evy} flights.
\newblock {\em Physical Review Letters}, 101(5):050601, August 2008.

\bibitem{majumdar_universal_2010}
Satya~N. Majumdar.
\newblock Universal first-passage properties of discrete-time random walks and
  {L\'evy} flights on a line: {Statistics} of the global maximum and records.
\newblock {\em Physica A: Statistical Mechanics and its Applications},
  389(20):4299--4316, October 2010.

\bibitem{nagy_statistical_2013}
Bela Nagy, J.~Doyne Farmer, Quan~M. Bui, and Jessika~E. Trancik.
\newblock Statistical basis for predicting technological progress.
\newblock {\em PLOS ONE}, 8(2):e52669, February 2013.

\bibitem{nevzorov_records_1988}
V.~B. Nevzorov.
\newblock Records.
\newblock {\em Theory of Probability \& Its Applications}, 32(2):201--228,
  January 1988.

\bibitem{nevzorov_record_1998}
Valery~B. Nevzorov and N.~Balakrishnan.
\newblock A record of records.
\newblock {\em Handbook of Statistics}, 16:515--570, January 1998.

\bibitem{noubary_tail_2010}
Reza~D Noubary.
\newblock Tail modeling, track and field records, and {Bolt}'s effect.
\newblock {\em Journal of Quantitative Analysis in Sports}, 6(3), 2010.

\bibitem{noubary_procedure_2005}
Reza~D Noubary.
\newblock A procedure for prediction of sports records.
\newblock {\em Journal of Quantitative Analysis in Sports}, 1(1), 2005.

\bibitem{redner_role_2006}
Sidney Redner and Mark~R. Petersen.
\newblock Role of global warming on the statistics of record-breaking
  temperatures.
\newblock {\em Physical Review E}, 74(6):061114, December 2006.

\bibitem{solow_how_2005}
Andrew~R Solow and Woollcott Smith.
\newblock How surprising is a new record?
\newblock {\em The American Statistician}, 59(2):153--155, May 2005.

\bibitem{terpstra_simple_2007}
Jeff~T Terpstra and Nicholas~D Schauer.
\newblock A simple random walk model for predicting track and field world
  records.
\newblock {\em Journal of Quantitative Analysis in Sports}, 3(3), 2007.

\bibitem{volf2010stochastic}
Petr Volf.
\newblock A stochastic model of progression of athletic records.
\newblock {\em IMA Journal of Management Mathematics}, page dpq010, 2010.

\bibitem{wergen_record_2012}
Gregor Wergen, Satya~N. Majumdar, and Gr\'egory Schehr.
\newblock Record statistics for multiple random walks.
\newblock {\em Physical Review E}, 86(1):011119, July 2012.

\bibitem{wergen2013records}
Gregor Wergen.
\newblock Records in stochastic processes—theory and applications.
\newblock {\em Journal of Physics A: Mathematical and Theoretical},
  46(22):223001, 2013.

\bibitem{yang_distribution_1975}
Mark C.~K. Yang.
\newblock On the distribution of the inter-record times in an increasing
  population.
\newblock {\em Journal of Applied Probability}, 12(1):148--154, 1975.

\bibitem{franke2010records}
Jasper Franke, Gregor Wergen, and Joachim Krug.
\newblock Records and sequences of records from random variables with a linear
  trend.
\newblock {\em Journal of Statistical Mechanics: Theory and Experiment},
  2010(10):P10013, 2010.

\bibitem{katz_power_1999}
J.~Sylvan Katz and Leon Katz.
\newblock Power laws and athletic performance.
\newblock {\em Journal of Sports Sciences}, 17(6):467--476, January 1999.

\bibitem{lippi_updates_2008}
Giuseppe Lippi, Giuseppe Banfi, Emmanuel~J. Favaloro, Joern Rittweger, and
  Nicola Maffulli.
\newblock Updates on improvement of human athletic performance: focus on world
  records in athletics.
\newblock {\em British Medical Bulletin}, 87(1):7--15, September 2008.

\bibitem{nevill_are_2005}
Alan~M. Nevill and Gregory Whyte.
\newblock Are there limits to running world records?
\newblock {\em Medicine and Science in Sports and Exercise}, 37(10):1785--1788,
  October 2005.

\bibitem{nevill_are_2007}
A.~M. Nevill, G.~P. Whyte, R.~L. Holder, and M.~Peyrebrune.
\newblock Are {There} {Limits} to {Swimming} {World} {Records}?
\newblock {\em International Journal of Sports Medicine}, 28(12):1012--1017,
  December 2007.

\bibitem{coumou2012decade}
Dim Coumou and Stefan Rahmstorf.
\newblock A decade of weather extremes.
\newblock {\em Nature climate change}, 2(7):491--496, 2012.

\bibitem{moore_cramming_1965}
G.E. Moore.
\newblock Cramming more components onto integrated circuits.
\newblock {\em Electronics}, 38(8):114--117, april 1965.

\bibitem{bettencourt_determinants_2013}
Lu\'is M.~A. Bettencourt, Jessika~E. Trancik, and Jasleen Kaur.
\newblock Determinants of the pace of global innovation in energy technologies.
\newblock {\em PLOS ONE}, 8(10):e67864, October 2013.

\bibitem{koh_functional_2006}
Heebyung Koh and Christopher~L. Magee.
\newblock A functional approach for studying technological progress:
  {Application} to information technology.
\newblock {\em Technological Forecasting and Social Change}, 73(9):1061--1083,
  November 2006.

\bibitem{mcnerney_role_2011}
James McNerney, J.~Doyne Farmer, Sidney Redner, and Jessika~E. Trancik.
\newblock Role of design complexity in technology improvement.
\newblock {\em Proceedings of the National Academy of Sciences},
  108(22):9008--9013, May 2011.

\bibitem{nagy_superexponential_2011}
Bela Nagy, J.~Doyne Farmer, Jessika~E. Trancik, and John~Paul Gonzales.
\newblock Superexponential long-term trends in information technology.
\newblock {\em Technological Forecasting and Social Change}, 78(8):1356--1364,
  October 2011.

\bibitem{lundstrom_moores_2003}
Mark Lundstrom.
\newblock Moore's {Law} forever?
\newblock {\em Science}, 299(5604):210--211, January 2003.

\bibitem{denny_limits_2008}
Mark~W. Denny.
\newblock Limits to running speed in dogs, horses and humans.
\newblock {\em Journal of Experimental Biology}, 211(24):3836--3849, December
  2008.

\bibitem{desgorces_oxford_2008}
Fran\c{c}ois-Denis Desgorces, Geoffroy Berthelot, Nour~El Helou, Val\'erie
  Thibault, Marion Guillaume, Muriel Tafflet, Olivier Hermine, and
  Jean-Fran\c{c}ois Toussaint.
\newblock From {Oxford} to {Hawaii} ecophysiological barriers limit human
  progression in ten sport monuments.
\newblock {\em PLOS ONE}, 3(11):e3653, November 2008.

\bibitem{prampero_factors_2003}
Pietro Enrico~di Prampero.
\newblock Factors limiting maximal performance in humans.
\newblock {\em European Journal of Applied Physiology}, 90(3-4):420--429,
  October 2003.

\bibitem{chang2011limit}
Yu~Sang Chang and Seung~Jin Baek.
\newblock Limit to improvement in running and swimming.
\newblock {\em International Journal of Applied Management Science},
  3(1):97--120, 2011.

\bibitem{el_helou_tour_2010}
Nour El~Helou, Geoffroy Berthelot, Val\'erie Thibault, Muriel Tafflet, Hala
  Nassif, Fr\'ed\'eric Campion, Olivier Hermine, and Jean-Fran\c{c}ois
  Toussaint.
\newblock Tour de {France}, {Giro}, {Vuelta}, and classic {European} races show
  a unique progression of road cycling speed in the last 20 years.
\newblock {\em Journal of Sports Sciences}, 28(7):789--796, May 2010.

\bibitem{neuts_waiting_1967}
Marcel~F. Neuts.
\newblock Waitingtimes between record observations.
\newblock {\em J. Appl. Probability}, 4:206--208, 1967.

\bibitem{feller1966book}
William Feller.
\newblock {\em An introduction to probability theory and its applications.
  {Vol.} {II}}.
\newblock John Wiley \& Sons, Inc., New York-London-Sydney, 1966.

\bibitem{ballerini_resnick_1985_linear}
Rocco Ballerini and Sidney Resnick.
\newblock Records from improving populations.
\newblock {\em J. Appl. Probab.}, 22(3):487--502, 1985.

\bibitem{ballerini_resnick_1987_linear}
Rocco Ballerini and Sidney~I. Resnick.
\newblock Records in the presence of a linear trend.
\newblock {\em Adv. in Appl. Probab.}, 19(4):801--828, 1987.

\bibitem{borovkov1999linear}
K.~Borovkov.
\newblock On records and related processes for sequences with trends.
\newblock {\em J. Appl. Probab.}, 36(3):668--681, 1999.

\bibitem{WFK2011linear}
Gregor Wergen, Jasper Franke, and Joachim Krug.
\newblock Correlations between record events in sequences of random variables
  with a linear trend.
\newblock {\em J. Stat. Phys.}, 144(6):1206--1222, 2011.

\bibitem{eilazar_klafter_2009_variance}
Iddo Eilazar and Joseph Klafter.
\newblock On the generation of anomalous diffusion.
\newblock {\em J. Phys. A: Math. Theory.}, 42:472003, 2009.

\bibitem{lenski1994dynamics}
Richard~E Lenski and Michael Travisano.
\newblock Dynamics of adaptation and diversification: a 10,000-generation
  experiment with bacterial populations.
\newblock {\em Proceedings of the National Academy of Sciences},
  91(15):6808--6814, 1994.

\bibitem{birnbaum_numerical_1952}
Z.~W. Birnbaum.
\newblock Numerical tabulation of the distribution of {K}olmogorov's statistic
  for finite sample size.
\newblock {\em J. Amer. Stat. Assoc.}, 47:425--441, 1952.

\bibitem{ely2007impact}
Matthew~R Ely, Samuel~N Cheuvront, William~O Roberts, and Scott~J Montain.
\newblock Impact of weather on marathon-running performance.
\newblock {\em Medicine and Science in Sports and Exercise}, 39(3):487--493,
  2007.

\bibitem{suping1992study}
Zhang Suping, Meng Guanglin, Wang Yanwen, and Li~Ji.
\newblock Study of the relationships between weather conditions and the
  marathon race, and of meteorotropic effects on distance runners.
\newblock {\em International journal of biometeorology}, 36(2):63--68, 1992.

\bibitem{anderson2011anderson}
Theodore~W Anderson.
\newblock Anderson--{Darling} tests of goodness-of-fit.
\newblock In {\em International Encyclopedia of Statistical Science}, pages
  52--54. Springer, 2011.

\end{thebibliography}

\end{document}